\documentclass[trackchanges,twocolumn,amsmath]{aastex701}

\usepackage{CJK}

\begin{document}
\begin{CJK*}{UTF8}{gkai}

\title{Funnel Leakage of the Wien Fireball in the Early Nebula after a Binary Neutron Star Merger}

\author[gname=Shao-Ze, sname=Li]{Shao-Ze Li (李少泽)}
\affiliation{College of Physics Science and Technology, Hebei University, Baoding 071002, People's Republic of China}
\affiliation{Hebei Key Laboratory of High-precision Computation and Application of Quantum Field Theory, Baoding 071002, People's Republic of China}
\affiliation{Hebei Research Center of the Basic Discipline for Computational Physics, Baoding 071002, People's Republic of China}
\email{lishaoze@mails.ccnu.edu.cn}
\correspondingauthor{lishaoze@mails.ccnu.edu.cn}

\author[gname=Yun-Wei, sname=Yu]{Yun-Wei Yu}
\affiliation{Institute of Astrophysics, Central China Normal University, Wuhan 430079, People's Republic of China}
\email{yuyw@mail.ccnu.edu.cn}

\author[gname=He, sname=Gao]{He Gao}
\affiliation{Institute for Frontier in Astronomy and Astrophysics, Beijing Normal University, Beijing 102206, People's Republic of China}
\affiliation{Department of Astronomy, Beijing Normal University, Beijing 100875, People's Republic of China}
\email{gaohe@bnu.edu.cn}

\author[gname=Peng-Cheng, sname=Yang]{Peng-Cheng Yang}
\affiliation{College of Physics Science and Technology, Hebei University, Baoding 071002, People's Republic of China}
\affiliation{Hebei Key Laboratory of High-precision Computation and Application of Quantum Field Theory, Baoding 071002, People's Republic of China}
\affiliation{Hebei Research Center of the Basic Discipline for Computational Physics, Baoding 071002, People's Republic of China}
\email{ypc32842@163.com}   
\author[gname=Yu, sname=Qu]{Yu Qu}
\affiliation{College of Physics Science and Technology, Hebei University, Baoding 071002, People's Republic of China}
\affiliation{Hebei Key Laboratory of High-precision Computation and Application of Quantum Field Theory, Baoding 071002, People's Republic of China}
\affiliation{Hebei Research Center of the Basic Discipline for Computational Physics, Baoding 071002, People's Republic of China}
\email{quyu5205354@163.com}   
\begin{abstract}
A newborn neutron star (NS) that survives after a binary NS merger can inject its rotational energy into the surrounding ejecta while the ejecta are still optically thick. The injected energy is redistributed by the pair cascade, and the nebula becomes a radiation-dominated plasma composed of photons and $e^\pm$ pairs. Compton scattering then governs the energy exchange while the photon number is conserved in scattering. This nebula is evolved with a one-zone model that includes pair creation, annihilation, photon injection, diffusion, and in particular the leakage through the jet-induced polar funnel. For typical parameters, we find that the Compton parameter $y$ is much larger than unity at early times. The nebula therefore remains in saturated Comptonization for 100 s to a few thousand seconds, depending on the magnetic field strength of the post-merger NS. The radiation field takes a Wien spectrum with a Compton equilibrium temperature. The nebula is then a so-called Wien fireball. The Compton temperature decreases from tens of keV to a few keV at early times. As long as the ejecta are optically thick, the funnel leakage is comparable to or exceeds the adiabatic loss. The funnel emission is quasi-thermal, and its isotropic-equivalent luminosity is much higher than the spin-down luminosity. We suggest that the quasi-thermal components observed in the extended emission of short gamma-ray bursts could come from the funnel leakage of the nebula fireball.
\end{abstract}


\section{Introduction}

Binary neutron star (BNS) mergers are multi-messenger sources that produce gravitational waves, short gamma-ray bursts (GRBs), and kilonovae. The joint detection of GW\,170817, GRB\,170817A, and AT\,2017gfo marks the beginning of the multi-messenger astronomy era \citep{abbott2017,cowperthwaite2017,drout2017,metzger2017welcome,tanvir2017}. After the merger, the remnant either survives as a massive neutron star (NS) or collapses to a black hole depending on the maximum NS mass $M_{\rm TOV}$, which is constrained by the NS equation of state \citep{margalitmetzger2017,aizhang2018,ai2020,rezzolla2018,shibatazhou2019}. A post-merger remnant can be a rapidly rotating, strongly magnetized NS (millisecond magnetar), in which dynamo action could amplify the magnetic field to as high as $B\sim10^{16}$ G \citep{price2006,zrake2013,kiuchi2024}. Near-Keplerian rotation ($P_{\rm i}\sim1$ ms) then gives an energy $E_{\rm rot}\sim10^{52}$--$10^{53}$ erg, released by spin-down and injected into the surrounding ejecta.

Spin-down energy injection from a newborn magnetar has long been invoked to explain GRB afterglow radiation \citep{dailu1998a,dailu1998b,zhangmeszaros2001}. The magnetar model was later applied to jet formation \citep{bucciantini2009,bucciantini2012,metzger2011proto}. Compared with long GRBs, short GRBs present a similar scenario, in which the early plateaus and extended emission can be powered by a millisecond massive magnetar formed after the binary NS merger, before it collapses \citep{zhang2013,lu2015,gibson2017,stratta2018,mosta2020}. Binary NS mergers also launch r-process-rich ejecta, which produce kilonova emission \citep{lipaczynski1998,metzger2010}. Searching for distinct bumps in short-GRB afterglow light curves has revealed kilonova candidates such as GRB\,130603B \citep{berger2013,tanvir2013}. If the massive NS formed after the merger survives long enough to inject its spin-down energy into the ejecta, it can produce a much brighter merger-nova \citep{fan2013magnetar,gao2013afterglow,gao2015mergernova,gao2017mergernova,yuzhanggao2013,yulidai2015,metzgerpiro2014}. For the GW\,170817 event, a long-lived NS can also account for the observed ``blue and red'' emission of AT\,2017gfo \citep{liliuyuzhang2018,radicelonglived2018,yuliu2017}. More recently, GRB\,211211A and GRB\,230307A have been identified as long-duration Type I GRBs that originate from compact binary mergers \citep{rastinejad2022,yang2024,zhang2025}. Both bursts last much longer than the $\sim2$\,s value that separates short from long GRBs. GRB\,211211A has a $\sim13$\,s main emission and a $\sim55$\,s extended emission, giving a total duration of $\sim68$\,s \citep{yang2022}, whereas GRB\,230307A has a prompt duration of $\sim42$\,s \citep{sun2025}. For GRB\,230307A, the emergence of a magnetar is inferred from its very early soft X-ray plateau \citep{zhong2024,sun2025}, and a brief $\sim$909\,Hz gamma-ray pulsation has been reported in the gamma-ray emission \citep{chen2025}. A long-lived magnetar has also been invoked to interpret the X-ray and potential kilonova emission of GRB\,211211A \citep{yang2022,fraija2026}. Alternatively, if the merger remnant collapses to a black hole, a strong magnetic flux threading the accretion flow can halt the inflow far from the innermost stable circular orbit and prolong the accretion lifetime \citep{gao2022}.

A particularly relevant clue comes from the extended-emission phases of GRB\,211211A and GRB\,230307A, the time-resolved spectra of which show quasi-thermal components with $kT\sim10$--$100$~keV \citep{chang2022,peng2024}. Independent analyses support a photospheric origin, including an evolving photosphere in GRB\,230307A \citep{wang2023photosphere,gupta2025} and photospheric emission modified by a radiation-mediated shock in GRB\,211211A \citep{wistemar2025}. Similar quasi-thermal components appear in the extended emission of other short GRBs \citep{chang2024}. These thermal signatures persist for tens to hundreds of seconds, much longer than the typical prompt emission ($T_{90}\lesssim2$~s) of short GRBs. For a BNS merger, sustaining a jet for tens to hundreds of seconds is difficult. Alternatively, NS--white dwarf (NS--WD) mergers have been proposed for Type I GRBs to explain both the anomalous long duration time and the association with kilonovae. The larger WD radius and longer tidal disruption or accretion timescale can sustain the central engine, while a newborn magnetar and $^{56}$Ni decay can power the following emission \citep{zhong2023nswd,chen2024nswd,wang2024nswd}.

Early kilonova models conventionally assume an isotropic ejecta, which simplifies estimates of the energy diffusion and the possible energy injection from the central NS. However, numerical simulations show that BNS merger ejecta are generically anisotropic as a natural consequence of the merger dynamics \citep[e.g.,][]{hotokezaka2013ejecta,sekiguchi2015,radice2016,shibatahotokezaka2019}. The anisotropic ejecta distribution also produces a diversity of kilonova emission \citep{gompertz2018diversity,kawaguchi2020diversity}. The ``blue and red' kilonova emission following GW\,170817 provides evidence for such an anisotropic ejecta distribution \citep{cowperthwaite2017,drout2017,tanvir2017}. In addition, the short GRB jet propagating through the ejecta can leave a relatively clean polar funnel after breakout; continued magnetar injection can help sustain this funnel and concentrate the injected energy near the poles \citep{wangzhangzhu2024}. At early times, the ejecta remain highly optically thick, so the NS energy is deposited inside or behind the ejecta. The pair cascade in the NS wind generates copious $e^\pm$ pairs. Synchrotron emission and inverse Compton scattering cool these pairs and populate the photon field, forming an $e^\pm$-photon plasma \citep{metzger2013supernova,metzgerpiro2014}. If such a funnel persists, $e^\pm$ pairs and photons can escape from the ejecta, producing observable polar funnel emission.

Therefore, this paper focuses on how the jet-induced funnel regulates energy transport and leakage from the nebula. At early times, around $t\sim100$~s, the $e^\pm$ and photon densities are extremely high, so Compton scattering dominates the process of energy exchange. The plasma in the early nebula reaches thermal equilibrium and shares a common temperature. With the photon number conserved in the Compton scattering, the resulting spectrum is a fully Comptonized Wien spectrum \citep{iwamototakahara2002,iwamototakahara2004}. Photons escaping through the funnel retain the Wien spectral distribution, which provides a possible source of the thermal component seen in the extended emission of a fraction of short GRBs. This paper is organized as follows. Section 2 discusses the physical properties of the early nebula and the Wien fireball (hereafter the nebula fireball); Section 3 presents energy diffusion and transport of the nebula fireball; Section 4 quantifies the leakage through the polar funnel and reveals the resulting observable signatures; Section 5 is the conclusion and discussion.

\section{Early Nebula Fireball}

A long-lived massive NS can survive after a BNS (or NS--WD) merger if the total mass of the remnant is below the maximum NS mass $M_{\rm TOV}$. Otherwise, as the total mass increases, the remnant can be a supramassive NS supported by uniform rotation, a hypermassive NS supported by differential rotation, or collapse promptly to a black hole \citep{sarin2021}. Such a surviving post-merger NS initially rotates near the Keplerian period \citep{shibatahotokezaka2019}. With a high field strength, magnetic dipole spin-down releases the rotational energy at a luminosity
\begin{equation}
\label{eq:Lsd}
L_{\rm sd}=L_0\left(1+\frac{t}{t_{\rm sd}}\right)^{-2},
\end{equation}
where $L_0=10^{47}\,R_{\rm s,6}^{6}\,B_{14}^{2}\,P_{\rm i,-3}^{-4}$\,erg\,s$^{-1}$ is the initial luminosity and $t_{\rm sd}=2\times10^{5}\,R_{\rm s,6}^{-6}\,B_{14}^{-2}\,P_{\rm i,-3}^{2}$\,s is the spin-down timescale \citep{dailu1998a,dailu1998b,zhangmeszaros2001}; $B$, $P_{\rm i}$, and $R_{\rm s}$ are the magnetic field strength, the initial spin period, and the NS radius, respectively. In this paper, $Q_{n}=Q/10^{n}$ is adopted in cgs units. For $B\sim10^{14}$--$10^{15}$\,G, the initial spin-down luminosity is $10^{47}$--$10^{49}$\,erg\,s$^{-1}$.

The injected energy is almost fully deposited as nonthermal or thermal radiation energy inside the nebula \citep{metzger2013supernova,metzgerpiro2014}, when the ejecta remain extremely optically thick. The energy density is $u=E/V\simeq L_{\rm sd}t/V$. Assuming a blackbody distribution for the photons, the blackbody temperature is
\begin{equation}
\label{eq:kTT}
\begin{aligned}
kT_{\rm bb}
&= k\left[\frac{3L_{\rm sd}}{4\pi a\,v^{3}t^{2}}\right]^{1/4}\\
&= 1.6\,{\rm keV}\,
L_{{\rm sd},48}^{1/4}\,v_{-1,c}^{-3/4}\,t_{2}^{-1/2},
\end{aligned}
\end{equation}
where $k$ is the Boltzmann constant, $a$ is the radiation constant, and $v$ is the expansion velocity of the ejecta. For $t\sim100$\,s, $L_{\rm sd}\sim10^{48}$\,erg\,s$^{-1}$, and $v\sim0.1c$, the photons fall in the keV range, and the spectrum peaks at $h\nu_{\rm bb}\sim2.8\,kT_{\rm bb}$. Such keV photons lie far below the threshold for pair production ($2m_ec^2\simeq1$\,MeV).
The NS wind initially consists of a Poynting flux and $e^\pm$ pairs, which can originate from the Goldreich--Julian current \citep{goldreichjulian1969}.
The pairs injected by the wind are relativistic. They cool through synchrotron radiation and inverse Compton (IC) scattering and lose their energy to high-energy photons. These photons easily exceed the threshold energy for pair creation. They then collide with the background photons of the nebula, creating new $e^\pm$ pairs. These pairs are again accelerated and cooled, entering a repeating cycle known as the pair cascade \citep{lightmanzdziarski1987,svensson1987,vurmpoutanen2009}. The cascade process is characterized by the compactness parameter,
\begin{equation}
\label{eq:ell}
\ell=\frac{E\,\sigma_T R}{V\,m_e c^2}.
\end{equation}
where $\sigma_T=6.65\times10^{-25}$\,cm$^2$ is the Thomson cross-section. At early times $\ell\gg1$, so the cascade is saturated \citep{svensson1987}. Following \citet{metzgerpiro2014}, the pair-creation rate is
\begin{equation}
\label{eq:Npmplus}
\dot N_\pm^{+}=\frac{Y L_{\rm sd}}{m_e c^2},
\end{equation}
where $Y\sim0.1$ is the pair multiplicity. Pair creation is accompanied by annihilation at the $e^+e^-$ collision rate,
\begin{equation}
\label{eq:Npmminus}
\dot N_\pm^{-}=\frac{3}{16}\,\sigma_T c\,N_\pm^{2}\,V^{-1}.
\end{equation}
Equating the creation and annihilation rates gives the equilibrium pair number,
\begin{equation}
\label{eq:Npm_eq}
N_\pm=\sqrt{\frac{16\,Y L_{\rm sd}V}{3\,\sigma_T\,m_e c^3}}.
\end{equation}
This large pair population raises the scattering optical depth of the nebula, which is estimated as
\begin{equation}
\label{eq:tau_n}
\begin{aligned}
\tau_{\mathrm n}&=\sigma_T R\,n_\pm\simeq\left(\frac{4Y L_{\rm sd}\sigma_T}{\pi R\,m_e c^3}\right)^{1/2}\\
&\simeq3.4\times10^{3}\,Y_{-1}^{1/2}\,L_{{\rm sd},48}^{1/2}\,v_{-1,c}^{-1/2}\,t_{2}^{-1/2},
\end{aligned}
\end{equation}
where $n_\pm=N_\pm/V$ is the pair number density. The large optical depth, $\tau_{\mathrm n}\gg1$, shows that the early nebula is optically thick and radiation-trapped, consistent with \citet{metzgerpiro2014}.

Besides creating copious $e^\pm$ pairs, the cascade converts the spin-down power into a power-law injection spectrum, $J_\nu\propto\nu^{-\alpha}$ with $\alpha\simeq1$ \citep[][a ``flat'' spectrum, $\nu\dot E_{{\rm sd},\nu}\simeq$ const]{svensson1987,metzger2013supernova}. The thermal photons of the nebula, Eq.\,\ref{eq:kTT}, serve as the low-energy bath of the cascade. The injection spectrum extends from the thermal bath to the pair-creation threshold,
\begin{equation}
\label{eq:Einjnu}
\dot E_{{\rm sd},\nu}\simeq\frac{L_{\rm sd}}{\Lambda\,\nu},
\qquad 3kT_{\rm bb}\lesssim h\nu\lesssim 2m_ec^2,
\end{equation}
where $\Lambda=\ln(h\nu_{\max}/h\nu_{\min})\simeq\ln[2m_ec^2/3kT_{\rm bb}]$. The injection spectrum is normalized by $\int\dot E_{{\rm sd},\nu}\,d\nu=L_{\rm sd}$. For the fiducial parameters of Eq.\,\ref{eq:kTT}, $kT_{\rm bb}\simeq1.6$\,keV, so $h\nu_{\min}\simeq3kT_{\rm bb}\simeq4.8$\,keV and $\Lambda\simeq\ln(2\times10^2)\simeq5.4$, i.e., the injected photons are mostly X-rays. The corresponding photon injection rate is
\begin{equation}
\label{eq:Nginj}
\begin{aligned}
\dot N_{\gamma,\rm inj}
&=\int_{\nu_{\min}}^{\nu_{\max}}\frac{\dot E_{{\rm sd},\nu}}{h\nu}\,d\nu \simeq 
\frac{L_{\rm sd}}{\Lambda\,(3kT_{\rm bb})}\\
&\simeq2.4\times10^{55}\,{\rm s^{-1}}\,L_{{\rm sd},48}
\left(\frac{kT_{\rm bb}}{1.6\,{\rm keV}}\right)^{-1}
\left(\frac{\Lambda}{5.4}\right)^{-1},
\end{aligned}
\end{equation}
The mean injected photon energy, $\langle h\nu\rangle\simeq L_{\rm sd}/\dot N_{\gamma,\rm inj}\simeq\Lambda\,(3kT_{\rm bb})\simeq26$\,keV, lies well below the pair-creation threshold $2m_ec^2$, so the injected photon number is carried by the low-energy end of the spectrum. As the fireball cools by adiabatic expansion, $h\nu_{\min}=3kT_{\rm bb}$ decreases and $\Lambda$ increases, so the number of injected photons per unit spin-down power rises.
The spectral state of the nebula at these early times is characterized by the Compton $y$ parameter. The nebula is extremely optically thick, and the continuous injection from the NS wind drives the pairs and photons into equilibrium under Compton heating and cooling, $T_e\simeq T_\gamma$. The Compton parameter is defined as
\begin{equation}
\label{eq:ycomp}
\begin{aligned}
y&=4\,\tau_{\mathrm n}\,\frac{kT_e}{m_ec^2}\\
&\simeq4.2\times10^{1}\,Y_{-1}^{1/2}\,L_{{\rm sd},48}^{1/2}\,v_{-1,c}^{-1/2}\,t_{2}^{-1/2}\,\left(\frac{kT_e}{1.6\,{\rm keV}}\right),
\end{aligned}
\end{equation}
For $y\ll1$ photons escape the nebula with negligible energy loss and the spectrum is nonthermal, whereas for $y\gg1$ they lose most of their energy before escaping, so the Comptonization saturates. The photons and pairs become strongly coupled by Compton scattering and reach a common temperature $T\simeq T_e\simeq T_\gamma$ instead of the blackbody temperature $T_{\rm bb}$. Compton scattering, however, only redistributes energy between photons and pairs. The photon number is conserved in Compton scattering, while double Compton scattering and bremsstrahlung are slow at this stage \citep{lightmanzdziarski1987}. The photon injection rate from the NS wind is high, but most of the injected photons are more energetic than those of the thermal bath. Therefore, the low-energy part of the blackbody spectrum cannot be filled efficiently. The fully thermalized early nebula develops a Wien spectrum with $\mu_\gamma<0$ rather than a blackbody spectrum with $\mu_\gamma=0$ \citep{iwamototakahara2002,iwamototakahara2004}. The Wien spectrum gives the photon number and energy densities,
\begin{equation}
\label{eq:Wien3}
\begin{aligned}
n_\gamma&\simeq\frac{16\pi}{(hc)^3}(kT)^3\,e^{\mu_\gamma/kT},\\
u_\gamma&\simeq\frac{48\pi}{(hc)^3}(kT)^4\,e^{\mu_\gamma/kT},
\end{aligned}
\end{equation}
so the mean photon energy and the peak of the energy spectrum are $\langle h\nu\rangle\simeq u_\gamma/n_\gamma\simeq3kT$ and $h\nu_{\rm pk}\simeq3kT$ (the number spectrum peaks at $2kT$). $T$ and $\mu_\gamma$ are not fixed by a blackbody assumption. The $kT_{\rm bb}$ of Eq.\,\ref{eq:kTT} is only an estimate of the trapped radiation energy from the blackbody relation, used to set the reference energy of the low-energy bath of the cascade. The actual $T$ and $\mu_\gamma$ are determined by the total photon number and energy of the photon field. Given $n_\gamma=N_\gamma/V$ and $u_\gamma=E_\gamma/V$, with $E_\gamma$ the total photon energy, one can obtain $T=T(N_\gamma,E_\gamma,V)$ and $\mu_\gamma=\mu_\gamma(N_\gamma,E_\gamma,V)$ from Eq.\,\ref{eq:Wien3}. Compared with a blackbody spectrum, a Wien spectrum at the same temperature contains fewer photons of higher mean energy, and this shortfall of photons relative to the energy is the signature of $\mu_\gamma<0$.

In saturated Comptonization the radiation field is in Wien equilibrium, so that Eq.\,\ref{eq:Wien3} gives $kT_{\rm C}\simeq u_\gamma/(3n_\gamma)\simeq E_\gamma/(3N_\gamma)$. Adopting $t\sim100$\,s and $L_{\rm sd}\sim10^{48}$\,erg\,s$^{-1}$ as reference values, the total photon energy is $E_\gamma\sim L_{\rm sd}t\sim10^{50}$\,erg, an upper bound that neglects the adiabatic loss. The photon number is dominated by cascade injection, $N_\gamma\sim\dot N_{\gamma,\rm inj}t\sim2.4\times10^{57}$. Substituting the above values gives,
\begin{equation}
\label{eq:TC}
\begin{aligned}
kT_{\rm C}&\simeq\frac{E_\gamma}{3N_\gamma}
\simeq\frac{L_{\rm sd}}{3\,\dot N_{\gamma,\rm inj}}\\
&\simeq\Lambda\,kT_{\rm bb}\simeq8.5\,{\rm keV}
\left(\frac{\Lambda}{5.4}\right)\left(\frac{kT_{\rm bb}}{1.6\,{\rm keV}}\right),
\end{aligned}
\end{equation}
so that $kT_{\rm C}$ is well above $kT_{\rm bb}$. The photon deficit therefore raises the temperature at a fixed energy density. Equation\,\ref{eq:Wien3} further gives $e^{\mu_\gamma/kT_{\rm C}}\sim1.3\times10^{-3}$, and $\mu_\gamma\sim-6.6\,kT_{\rm C}$, which is self-consistent with the Wien limit.

Taken together, these estimates show that the early nebula at $t\sim100$\,s is a trapped $e^\pm$--photon plasma injected by the NS wind. It is radiation-dominated, extremely optically thick ($\tau_{\mathrm n}\gg1$), and in saturated Comptonization ($y\gg1$). In this sense, the early nebula is a Wien fireball, with the radiation field in Wien equilibrium and $\mu_\gamma<0$ \citep{iwamototakahara2002,iwamototakahara2004}. Similar Comptonized fireballs appear in the models and spectral fits of \citet{zhang2023compton} and \citet{yin2024compton}. An adiabatically expanding nebula fireball loses energy rapidly through adiabatic work. In the early nebula, however, the optically thick ejecta confine the injected energy to the fireball interior. The competition between adiabatic expansion and injection from the NS maintains the fireball temperature at the keV level.

\section{Fireball Evolution}

At early times both the ejecta and the nebula fireball are extremely optically thick, so the energy injected by the NS wind is almost fully deposited and thermalized. The thermal photons diffuse outward by a random walk and are repeatedly scattered by the $e^\pm$ pairs of the nebula and the ejecta material \citep{kasenbildsten2010,yuzhanggao2013,yulidai2015}. For an isotropic ejecta the diffusion luminosity can be estimated as
\begin{equation}
\label{eq:Le}
L_{\rm e}\simeq\frac{E}{t_{\rm d,n}+t_{\rm d,e}}.
\end{equation}
The photon diffusion timescales of the nebula fireball and the ejecta are
\begin{equation}
\label{eq:tdn}
t_{\rm d,n}=\begin{cases} \tau_{\mathrm n} R_{\mathrm n}/c, & \tau_{\mathrm n}>1,\\ R_{\mathrm n}/c, & \tau_{\mathrm n}<1,\end{cases}
\end{equation}
\begin{equation}
\label{eq:tde}
t_{\rm d,e}=\begin{cases} \tau_{\mathrm e} R_{\mathrm e}/c, & \tau_{\mathrm e}>1,\\ R_{\mathrm e}/c, & \tau_{\mathrm e}<1,\end{cases}
\end{equation}
where $\tau_{\mathrm e}\simeq3\kappa_e M_{\rm ej}/(4\pi R_{\mathrm e}^2)$ is the optical depth of the ejecta, $\kappa_e$ is its opacity, $R_{\mathrm n}\simeq R_{\mathrm e}\simeq v\,t$ are the radii of the nebula and the ejecta, and $E$ is the total energy deposited in the fireball. With radiation pressure dominating, roughly $E\simeq E_\gamma$. For the overall diffusion timescale, $t_{\rm d}=t_{\rm d,n}+t_{\rm d,e}$ is adopted.

Besides diffusion, the nebula fireball loses energy through adiabatic expansion,
\begin{equation}
\label{eq:LPdV}
L_{PdV}=P\,\frac{dV}{dt}=\frac{E}{3V}\,\frac{dV}{dt},
\end{equation}
where $P=E/(3V)$ is the radiation pressure. The timescale for adiabatic loss is $t_{PdV}=E/L_{PdV}=R_{\mathrm e}/v$, which is roughly the dynamical timescale. The ratio to the diffusion luminosity at early times is
\begin{equation}
\frac{L_{PdV}}{L_{\rm e}}\simeq(\tau_{\mathrm n}+\tau_{\mathrm e})\,\frac{v}{c},
\end{equation}
which is much larger than unity at early times. Adiabatic work therefore dominates the energy loss. Injection and adiabatic loss keep the temperature at the keV level until the optical depth declines and diffusion takes over the energy loss at $t_{\rm d}\sim R/v$.

The fireball is described by the following variables, the volume $V$, the total energy $E$, the pair number $N_\pm$, the photon number $N_\gamma$, and the ejecta velocity $v$. The volume follows $V=4\pi R^3/3$, and the radius follows $dR/dt=v$. The total energy evolves as \citep{kasenbildsten2010,yuzhanggao2013,yulidai2015}\footnote{Here we neglect the absorption and reflection of photons by the ejecta, as in \citet{metzger2013supernova} and \citet{metzgerpiro2014}. The timescale for the absorbed photons to ionize the ejecta is several hours, which is much longer than the timescale considered in this paper. At early times $\tau_{\rm n}\gg1$, so most photons are trapped inside the nebula and only a small fraction diffuses outward into the ejecta. The absorption and ionization by the ejecta, and the bulk diffusion, are actually negligible compared with the adiabatic work at early times.}
\begin{equation}
\label{eq:Etot}
\frac{dE}{dt}=L_{\rm sd}-P\,\frac{dV}{dt}-L_{\rm e},
\end{equation}
where the first term is the injection from the NS, the second is the adiabatic loss, and the third is the diffusion loss. At early times the pair and photon numbers are governed by injection and annihilation. The pair number evolves as
\begin{equation}
\label{eq:Npm}
\frac{dN_\pm}{dt}=\frac{Y\,L_{\rm sd}}{m_e c^2}-\frac{3}{16}\,\sigma_T c\,\frac{N_\pm^2}{V},
\end{equation}
where the first term is the cascade injection and the second term is non-relativistic annihilation, with $\langle\sigma_{\rm ann} v\rangle\simeq3\sigma_T c/16$ for $kT\ll m_ec^2$ \citep{metzgerpiro2014}. The photon number evolves together with the pair number,
\begin{equation}
\label{eq:Ng}
\frac{dN_\gamma}{dt}=\dot N_{\gamma,\rm inj}+2\times\frac{3}{16}\,\sigma_T c\,\frac{N_\pm^2}{V}-\frac{N_\gamma}{t_{\rm d}},
\end{equation}
where the first term is the injection from the cascade, the second is released in the $e^\pm$ annihilation, and the third is due to the diffusion loss. At early times, when $\tau_{\mathrm e}\gg1$, the diffusion loss can be neglected. The injection spectrum of Eq.\,\ref{eq:Einjnu} is the net photon output of the saturated cascade, with the low-energy cutoff at the thermal bath $3kT_{\rm bb}$. It is truncated at the pair-creation threshold $2m_ec^2$, so the photons consumed by pair creation are already accounted for in the cascade spectrum. No separate pair-creation term is therefore added to Eq.\,\ref{eq:Ng}. The radiation pressure accelerates the ejecta,
\begin{equation}
\label{eq:dvdt}
\frac{dv}{dt}=\frac{P}{M_{\rm ej}\,v}\,\frac{dV}{dt},
\end{equation}
where $M_{\rm ej}$ is the ejecta mass. The temperature $T$ and the photon chemical potential $\mu_\gamma$ are obtained from the spectral relation of Eq.\,\ref{eq:Wien3}.

\begin{figure}[t]
\centering
\includegraphics[width=\columnwidth]{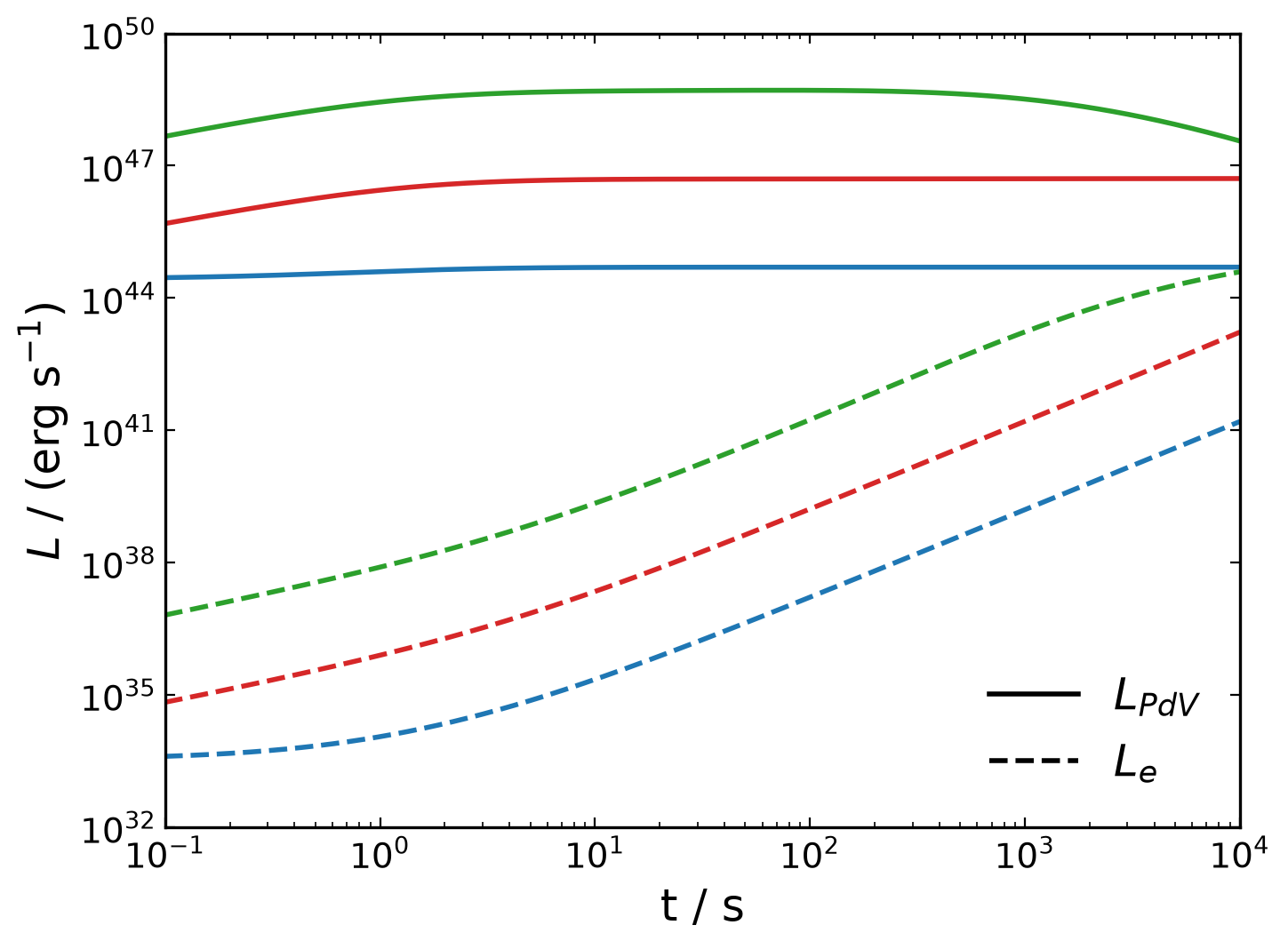}
\caption{Adiabatic loss power and diffusion luminosity, for $P=1$\,ms and $M_{\rm ej}=0.06\,M_\odot$. Blue, red, and green lines correspond to $B=10^{13}$, $10^{14}$, and $10^{15}$\,G. The adiabatic loss always exceeds the diffusion luminosity at early times, so the energy loss is dominated by adiabatic work.}
\label{fig:LPdV}
\end{figure}

\begin{figure}[t]
\centering
\includegraphics[width=\columnwidth]{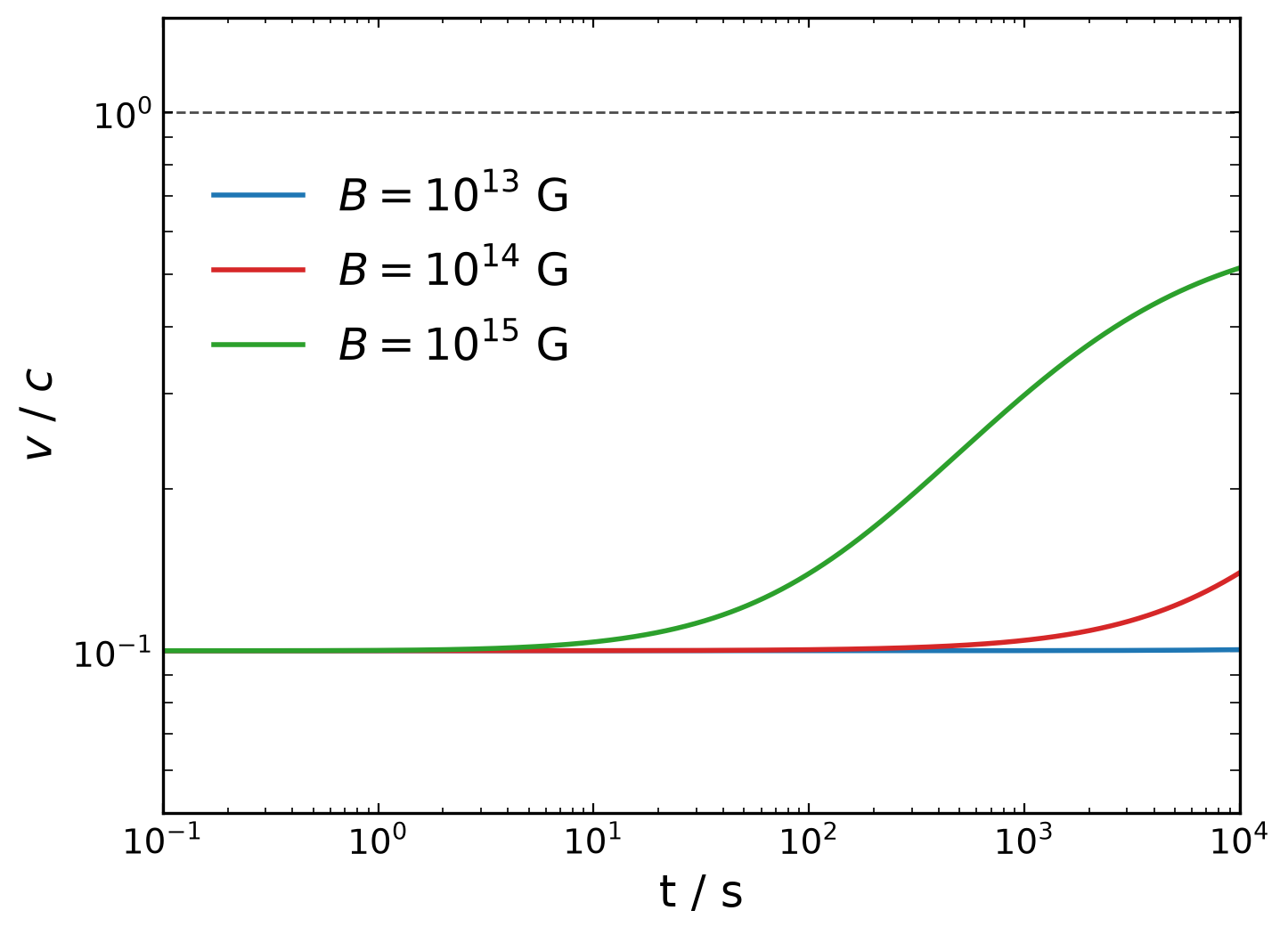}
\caption{Velocity of the ejecta, in units of $c$, for $P=1$\,ms and $M_{\rm ej}=0.06\,M_\odot$. Blue, red, and green lines correspond to $B=10^{13}$, $10^{14}$, and $10^{15}$\,G. The horizontal dashed line marks $v=c$. The ejecta remain sub-relativistic at early times.}
\label{fig:v}
\end{figure}

\begin{figure}[t]
\centering
\includegraphics[width=\columnwidth]{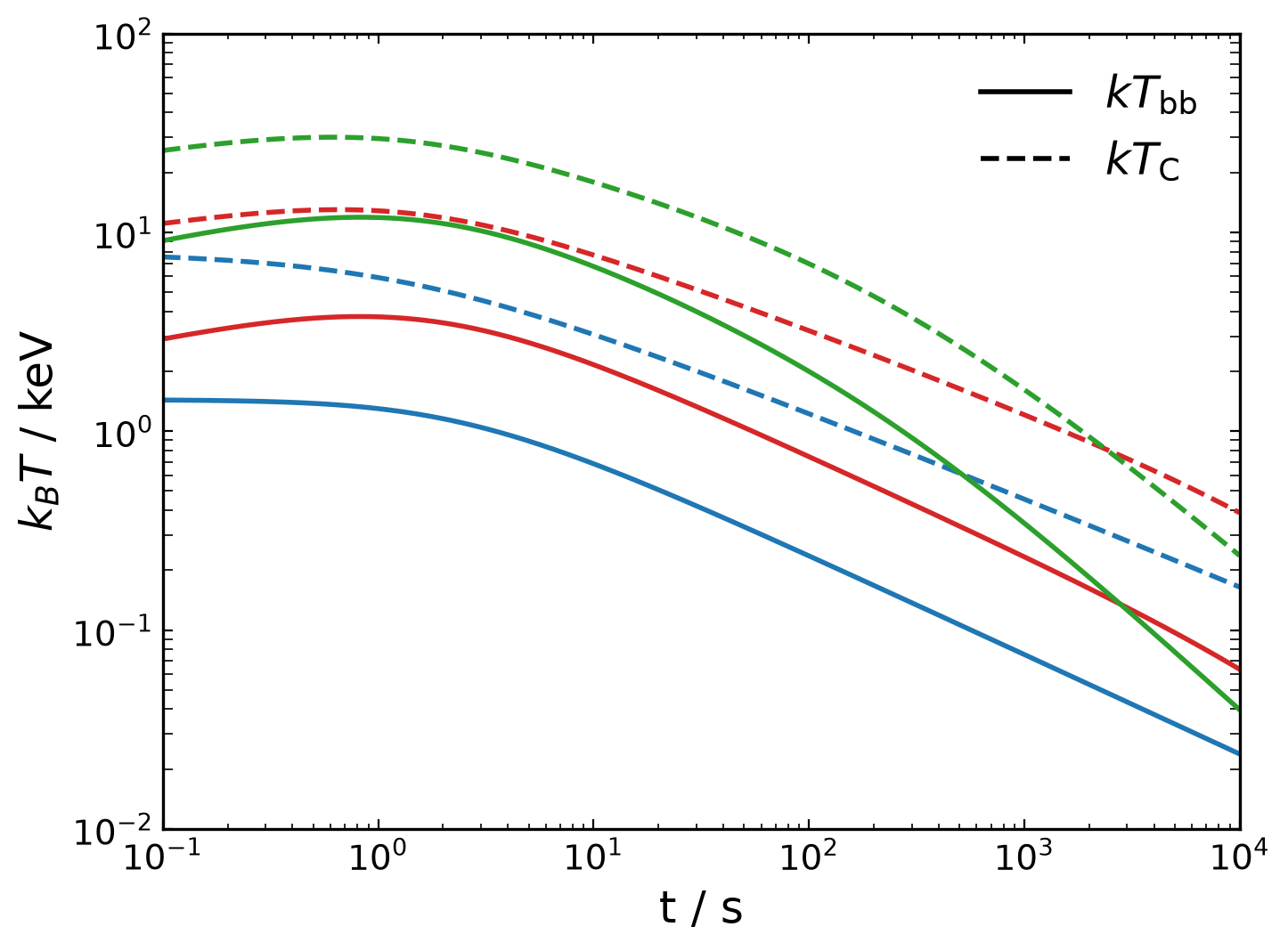}
\caption{Blackbody temperature, $kT_{\rm bb}=(E/(aV))^{1/4}$, and Wien (Compton) temperature, $kT_{\rm C}=E/(3N_\gamma)$. Blue, red, and green lines correspond to $B=10^{13}$, $10^{14}$, and $10^{15}$\,G. The parameters are the same as in Fig.\,\ref{fig:LPdV}.}
\label{fig:kT}
\end{figure}

\begin{figure}[t]
\centering
\includegraphics[width=\columnwidth]{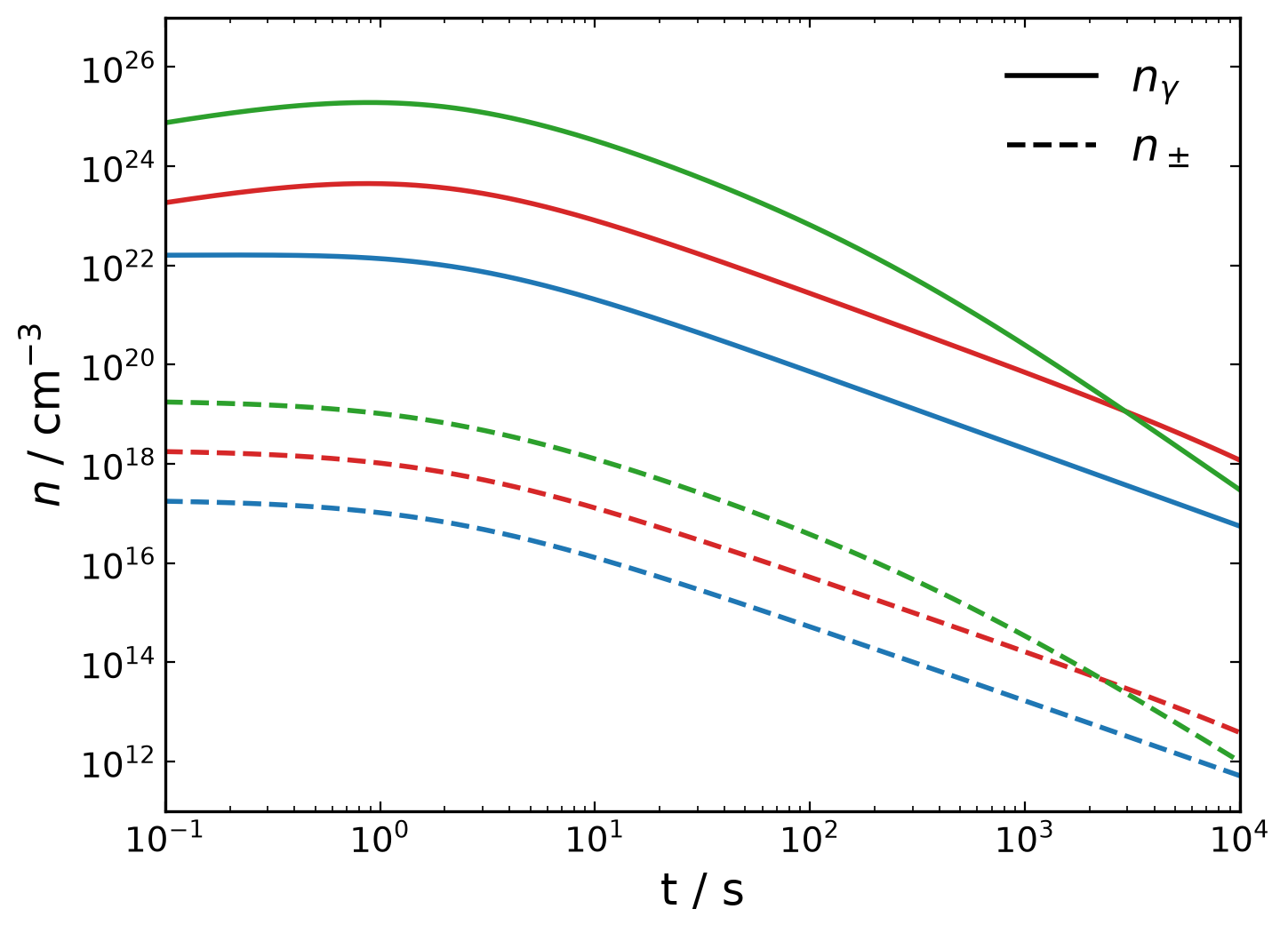}
\caption{Photon density $n_\gamma$ and pair density $n_\pm$. Blue, red, and green lines correspond to $B=10^{13}$, $10^{14}$, and $10^{15}$\,G. The parameters are the same as in Fig.\,\ref{fig:LPdV}.}
\label{fig:ndens}
\end{figure}

The typical ejecta parameters are adopted as $M_{\rm ej}=0.06\,M_\odot$ and $\kappa_e=1$\,cm$^2$\,g$^{-1}$ \citep{metzger2020kilonova}. The NS has an initial spin period $P_0=1$\,ms, and magnetic field strengths $B_0=10^{13}$, $10^{14}$, and $10^{15}$\,G are considered. Fig.\,\ref{fig:LPdV} shows the two channels for energy loss. The adiabatic loss significantly exceeds the diffusion loss at early times. The ratio between them decreases as the optical depth drops. It is still $\sim10^{3}$ at $t=10^{4}$\,s, so diffusion has not yet taken over the energy loss at early times. The ejecta are accelerated but still remain sub-relativistic, as Fig.\,\ref{fig:v} shows. The ejecta mass is set to $M_{\rm ej}=0.06\,M_\odot$ following the GW\,170817 event, and the rotational energy of the NS is not sufficient to accelerate such an ejecta to relativistic speeds. For a much smaller ejecta mass of $\sim10^{-4}\,M_\odot$, the ejecta could be accelerated to relativistic velocities. Fig.\,\ref{fig:kT} compares the two temperatures of the fireball. The Compton temperature always exceeds the blackbody temperature by a factor of $\sim$3--7. The fireball is therefore photon-starved, with $\mu_\gamma<0$. Fig.\,\ref{fig:ndens} shows the corresponding densities. At $t\sim100$\,s the photon density $n_\gamma=N_\gamma/V$ is $\sim10^{20}$--$10^{23}$\,cm$^{-3}$ over the three field strengths, and the pair density $n_\pm=N_\pm/V$ is $\sim10^{15}$--$10^{17}$\,cm$^{-3}$. Throughout, $n_\pm\lambda_{\rm dB}^3\ll1$, where $\lambda_{\rm dB}=h/(2\pi m_e kT)^{1/2}$ is the thermal de Broglie wavelength of the pairs, so the pairs remain non-degenerate.

\begin{figure}[t]
\centering
\includegraphics[width=\columnwidth]{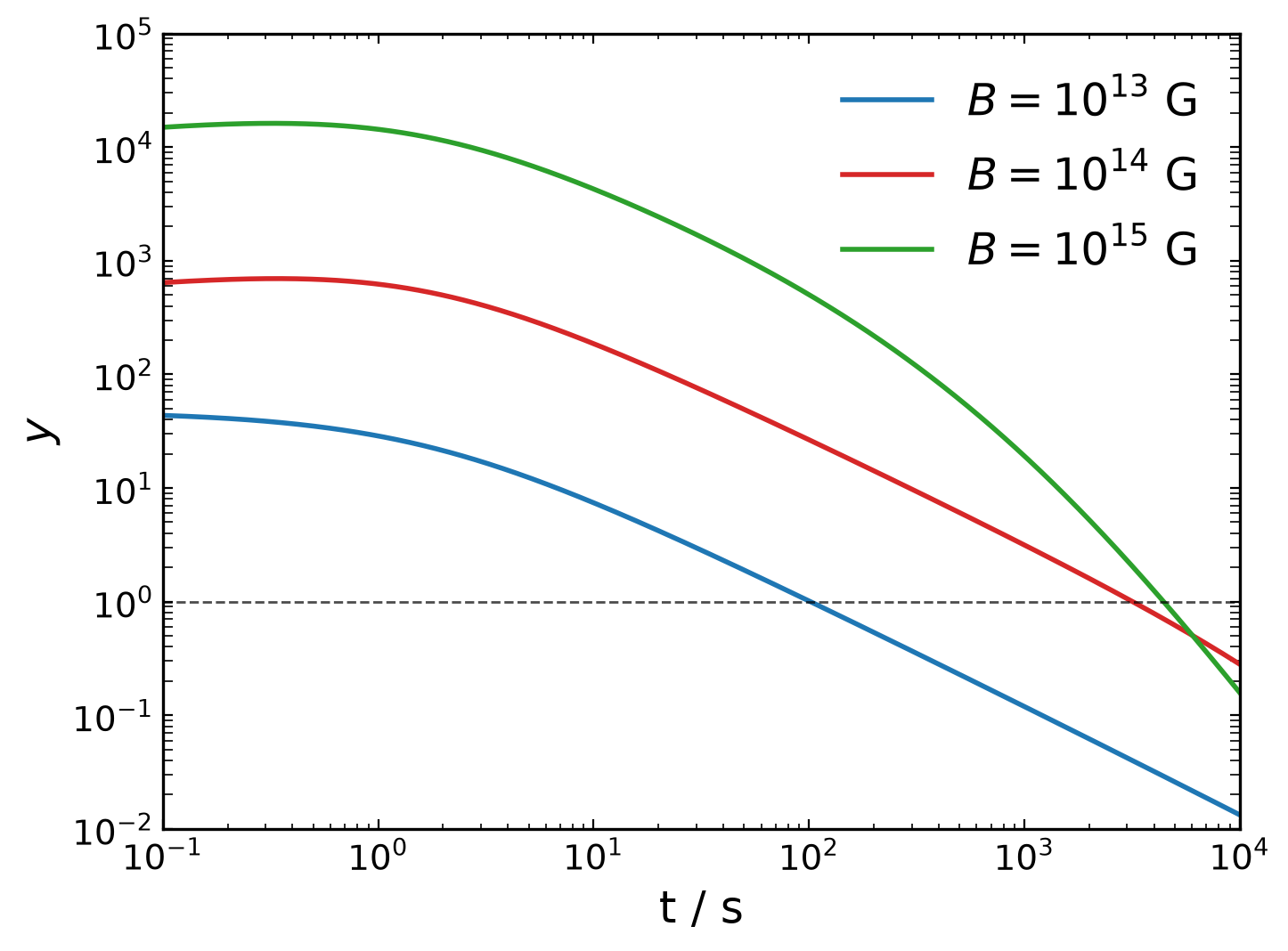}
\caption{Compton parameter $y$ (the horizontal dashed line marks $y=1$). At early times $y\gg1$, corresponding to saturated Comptonization and a Wien spectrum. The parameters are the same as in Fig.\,\ref{fig:LPdV}.}
\label{fig:ycomp}
\end{figure}

The spectral state of the fireball is further characterized by the Compton $y$ parameter. Fig.\,\ref{fig:ycomp} shows that $y$ decreases with time. At early times, $y$ increases with the magnetic field strength of the NS. At $t\sim100$\,s all three field strengths have $y\ge1$. Then the photons and pairs are fully Comptonized and share a common temperature. As the fireball expands, both $\tau_{\mathrm n}$ and the temperature decrease. The parameter $y$ eventually falls below unity. At that time, Comptonization is no longer saturated. The spectrum relaxes from the Wien form to the flat nonthermal injection spectrum of the cascade, and the fireball enters the nonthermal stage described by the Metzger et al. models \citep{metzger2013supernova,metzgerpiro2014}. The time $y\sim1$ therefore marks the end of the nebula fireball phase.

The results above assume a homogeneous, isotropic ejecta, that is, a spherically symmetric one-zone model. The fireball remains buried inside the ejecta due to the extremely large optical depth. Any radiation that reaches the surface has traversed the whole optically thick ejecta and has been absorbed and re-emitted by the ejecta \citep{metzgerpiro2014}, losing the information about the nebula fireball.

\section{Funnel Leakage}

BNS mergers are generally expected to launch a relativistic jet along the polar direction. If the jet successfully breaks out of the ejecta, it produces a short GRB associated with a kilonova \citep{abbott2017}. As the jet propagates through the ejecta it inflates a cocoon at the breakout opening (e.g., \citealp{nakarpiran2017}) and leaves a relatively clean funnel along the jet trajectory. Continued injection from the NS works against the refilling by the surrounding ejecta, so that such a jet-induced funnel can persist for a duration of time. Within the nebula fireball the $e^\pm$ pairs and the photons are tightly coupled by Compton scattering. The fireball energy is mostly deposited in the photons rather than in the pairs, including the rest energy $m_ec^2$ and the kinetic energy of the pairs. Radiation energy is rapidly redistributed by advection in the plasma, and the pressure is nearly isotropic, so the energy injected by the NS wind is shared over the whole fireball instead of remaining along the injection direction. The fireball energy can therefore flow into the funnel and escape, rather than diffuse through the ejecta or do adiabatic work. An observer near the funnel direction would then receive radiation from the fireball interior. Relativistic magnetohydrodynamic simulations also support that most of the injected energy is lost along the jet axis \citep{wangzhangzhu2024}.

For a standard bipolar jet or funnel of half-opening angle $\theta$, the two funnels cover a solid angle $\Omega=4\pi(1-\cos\theta)$. The beaming factor is
\begin{equation}
\label{eq:fbeam}
f_{\rm b}=\frac{\Omega}{4\pi}=1-\cos\theta.
\end{equation}
It sets the angular fraction of the fireball energy that can escape along the polar direction.
The funnel leakage consists of two parts. One is the diffusion and escape of photons through the nebula along the funnel. The other is the advective outflow of the $e^\pm$--photon plasma along the funnel. Both are confined to the funnel, which covers the solid-angle fraction $f_{\rm b}$, while the remaining fraction $1-f_{\rm b}$ corresponds to the ejecta direction. The total funnel leakage is
\begin{equation}
\label{eq:Lfun}
L_{\rm fun}=L_{\rm fd}+L_{\rm fo},
\end{equation}
where the first is the funnel diffusion and the second is the funnel outflow.
Photons can diffuse through the nebula $e^\pm$ layer to the funnel and escape without crossing the outer ejecta. The luminosity of funnel diffusion is
\begin{equation}
\label{eq:Lfd}
L_{\rm fd}=f_{\rm b}\,\frac{E}{t_{{\rm d},{\mathrm n}}}.
\end{equation}
Along the funnel direction the photons diffuse only through the nebula, so the diffusion timescale is $t_{{\rm d},{\mathrm n}}$. Because only the solid-angle fraction $f_{\rm b}$ is open, the corresponding diffusion-loss timescale is $t_{{\rm d},{\mathrm n}}/f_{\rm b}$. For the ejecta direction the diffusion timescale is $t_{{\rm d},{\mathrm n}}+t_{{\rm d},{\mathrm e}}$. If this diffusion-loss timescale is shorter than the total diffusion timescale $t_{{\rm d},{\mathrm n}}+t_{{\rm d},{\mathrm e}}$, energy diffuses out more efficiently along the funnel direction than along the ejecta direction.
Compared with funnel diffusion, funnel outflow plays a more significant role in the energy loss along the funnel direction. The pairs that fill the nebula are supplied by the cascade, and their density stays high while the Wien fireball exists ($\tau_{\mathrm n}\gg1$). In that regime photons and pairs stay coupled. The radiation pressure then drives a relativistic $e^\pm$--photon outflow along the funnel direction. If the ambient pressure outside the funnel is far below the fireball pressure, the outflow is choked and its velocity is limited to the local sound speed of the shocked material \citep{begelman1989,irwin2019}. With $\beta_{\rm s}=c_{\rm s}/c=1/\sqrt3$ and $\Gamma_{\rm s}=(1-\beta_{\rm s}^{2})^{-1/2}=\sqrt{3/2}$, the energy outflow rate is
\begin{equation}
\label{eq:Lfo}
L_{\rm fo}\simeq\frac{4}{3}\,u_\gamma\,\Gamma_{\rm s}^2\beta_{\rm s} c\,f_{\rm b}\,4\pi R^2
\simeq\zeta\,f_{\rm b}\,\frac{E}{R}\,c,
\end{equation}
where $\zeta=4\Gamma_{\rm s}^2\beta_{\rm s}\simeq3.5$. The choking requires the solid angle of the funnel to be much smaller than $4\pi$, i.e., $f_{\rm b}\ll1$. If this condition is not well satisfied, Eq.\,\ref{eq:Lfo} can be taken as an upper limit for the funnel outflow.
Comparing these two terms,
\begin{equation}
\label{eq:Lratio}
\frac{L_{\rm fo}}{L_{\rm fd}}\simeq\zeta\,\frac{c\,t_{{\rm d},{\mathrm n}}}{R}\simeq\zeta\,\tau_{\mathrm n}\gg1,
\end{equation}
where $t_{{\rm d},{\mathrm n}}=\tau_{\mathrm n}R/c$ in this regime. At early times, $\tau_{\mathrm n}\gg1$, the funnel outflow therefore dominates the leakage. As $\tau_{\mathrm n}$ decreases, the ratio falls toward $\zeta$, and the two terms become comparable at $\tau_{\mathrm n}\sim1/\zeta$. Beyond that point the leakage becomes dominated by the funnel diffusion, and the system enters the normal nebula phase with a nonthermal photon spectrum.

With the funnel leakage, the energy equation (Eq.\,\ref{eq:Etot}) is changed to
\begin{equation}
\label{eq:Efun}
\frac{dE}{dt}=L_{\rm sd}-(1-f_{\rm b})\Big(P\,\frac{dV}{dt}+L_{\rm e}\Big)-L_{\rm fun}.
\end{equation}
In this construction the adiabatic work and the diffusion $L_{\rm e}$ act only on the ejecta direction. Since $L_{\rm fo}\gg L_{\rm fd}$ at early times, the leakage is dominated by the funnel outflow. Therefore, photons and pairs leave the fireball at the same fractional rate, as expected for a tightly coupled plasma. With $t_{\rm fun}=E/L_{\rm fun}$, Eqs.\,\ref{eq:Npm} and \ref{eq:Ng} become
\begin{equation}
\label{eq:Npmfun}
\frac{dN_\pm}{dt}=\frac{Y L_{\rm sd}}{m_ec^2}-\frac{3}{16}\sigma_Tc\,\frac{N_\pm^2}{V}-\frac{N_\pm}{t_{\rm fun}},
\end{equation}
\begin{equation}
\label{eq:Ngfun}
\frac{dN_\gamma}{dt}=\dot N_{\gamma,\rm inj}+2\,\frac{3}{16}\sigma_Tc\,\frac{N_\pm^2}{V}-\frac{(1-f_{\rm b})N_\gamma}{t_{\rm d}}-\frac{N_\gamma}{t_{\rm fun}},
\end{equation}
where each escaping photon or pair carries away on average an energy $\sim E/(N_\gamma+N_\pm)\simeq3kT_{\rm C}$.

\begin{figure}[t]
\centering
\includegraphics[width=\columnwidth]{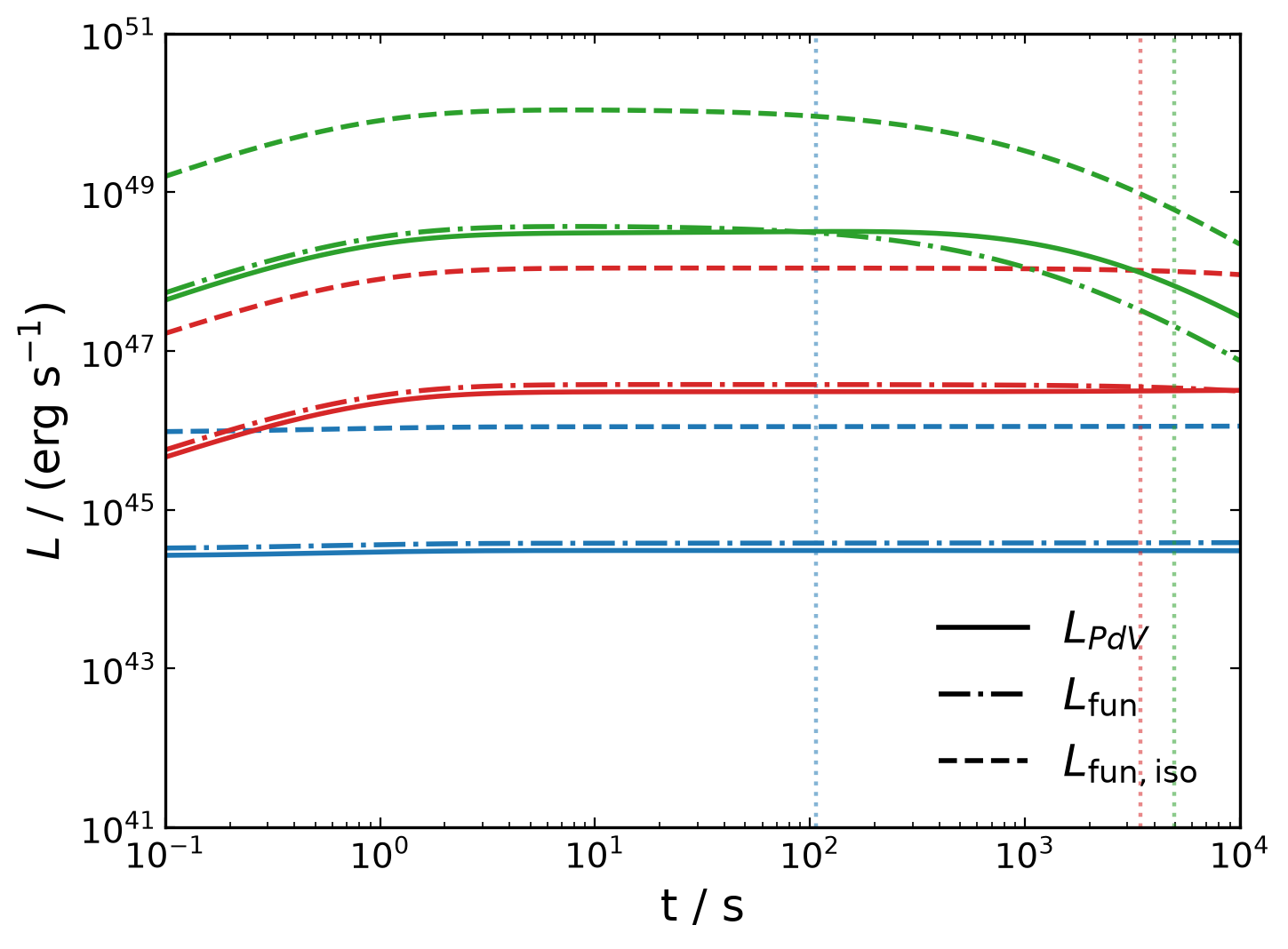}
\caption{Adiabatic loss power, total funnel leakage $L_{\rm fun}$, and isotropic-equivalent luminosity $L_{{\rm fun,iso}}=L_{\rm fun}/f_{\rm b}$, for $\theta=15^\circ$ and $P=1$ ms. Blue, red, and green lines correspond to $B=10^{13}$, $10^{14}$, and $10^{15}$\,G. The vertical dotted lines of the same colors mark when the Compton $y$ parameter reaches unity for each field.}
\label{fig:fLPdV}
\end{figure}

\begin{figure}[t]
\centering
\includegraphics[width=\columnwidth]{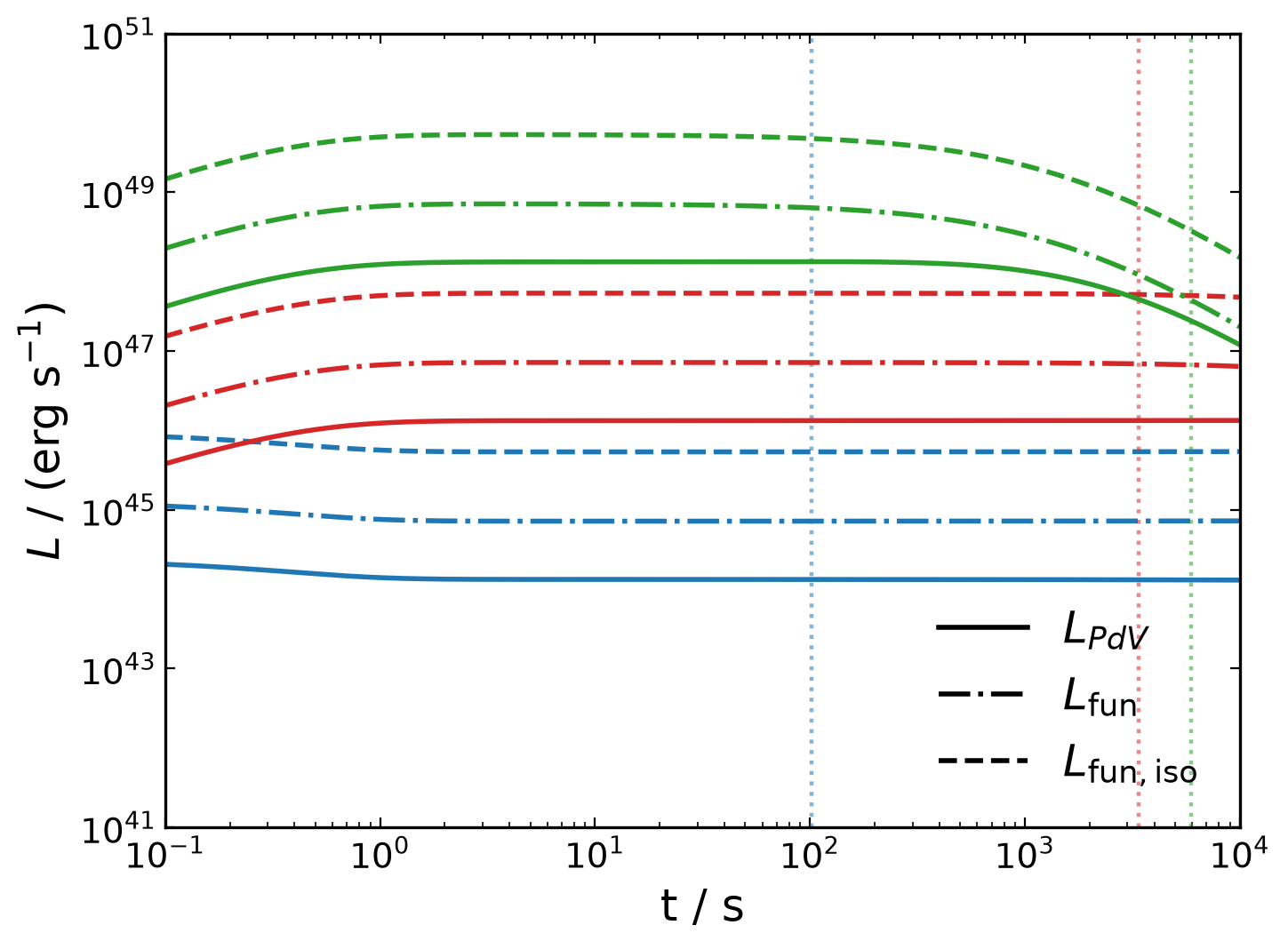}
\caption{The same as in Fig.\,\ref{fig:fLPdV}, but for $\theta=30^\circ$.}
\label{fig:fLPdV30}
\end{figure}

\begin{figure}[t]
\centering
\includegraphics[width=\columnwidth]{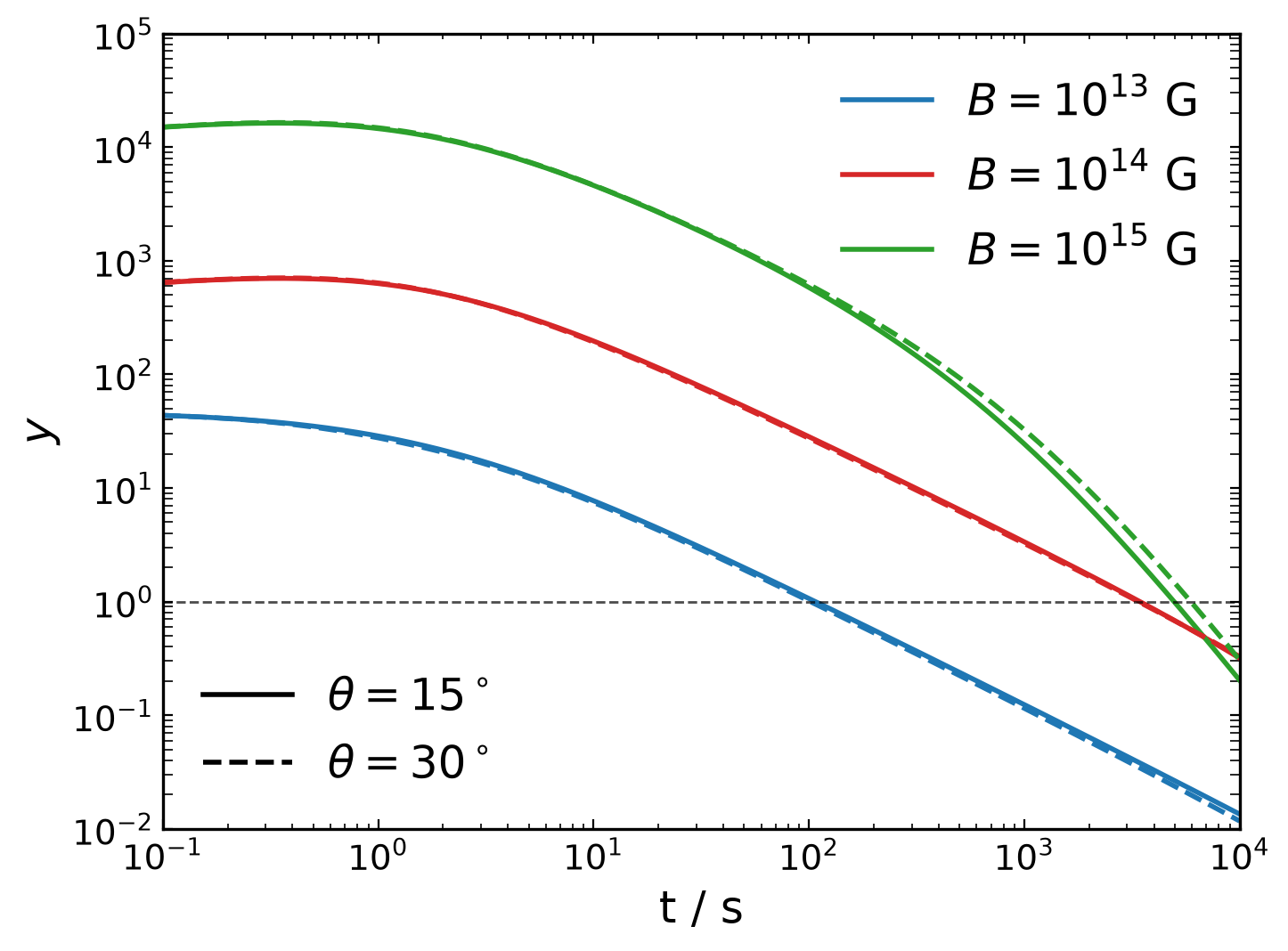}
\caption{Compton parameter $y$ as a function of time when the funnel leakage is included, for $P=1$ ms. Solid lines correspond to $\theta=15^\circ$ and dashed lines to $\theta=30^\circ$. The funnel shifts the time at which $y$ reaches unity to the right by at most about $30\%$ compared to the no-funnel case of Fig.\,\ref{fig:ycomp}.}
\label{fig:ycompfunnel}
\end{figure}

Equations\,\ref{eq:Efun}--\ref{eq:Ngfun} are integrated numerically for $\theta=15^\circ$ and $30^\circ$ ($f_{\rm b}\simeq0.034$ and $0.134$), for which the choking condition is satisfied. The other parameters are the same as in Section 3. Figs.\,\ref{fig:fLPdV} and\,\ref{fig:fLPdV30} show the light curves of different luminosities. In both, the funnel leakage $L_{\rm fun}$ is higher than the adiabatic loss $L_{PdV}=(1-f_{\rm b})Ev/R$ at early times. As $f_{\rm b}$ increases, the funnel area grows and $L_{\rm fun}$ becomes higher, while the adiabatic work is reduced. This is because a larger $f_{\rm b}$ allows more energy to leak through the funnel. If the leaked energy always continues along the funnel direction, an observer inside the beam would infer an isotropic-equivalent luminosity of approximately
\begin{equation}
\label{eq:Liso}
L_{{\rm fun,iso}}\simeq\frac{L_{\rm fun}}{f_{\rm b}},
\end{equation}
shown as dashed lines in Figs.\,\ref{fig:fLPdV} and\,\ref{fig:fLPdV30}. In contrast, $L_{{\rm fun,iso}}$ is higher by a factor of $\sim$2 for $\theta=15^\circ$ than for $\theta=30^\circ$. The actual leakage rate is set by $L_{\rm fun}$, so the more slowly the funnel leaks, the longer the fireball persists and the higher the isotropic-equivalent luminosity inferred along the funnel direction. The pair escape term $N_\pm/t_{\rm fun}$ remains small relative to the cascade injection rate, so $N_\pm$ (and hence $\tau_{\mathrm n}$) is still governed by injection and annihilation. The evolution of the Compton $y$ parameter therefore stays close to the case without the funnel, as shown in Fig.\,\ref{fig:ycompfunnel}.
Although choking limits the acceleration, the pair-photon outflow accelerates further by adiabatic expansion once it has passed through the funnel. Whether the energy escapes freely or is converted into bulk kinetic energy is set by the optical depth of the outflow itself rather than by that of the whole nebula.
With a coefficient $\eta$ describing the acceleration, the outflow may reach a mild Lorentz factor
\begin{equation}
\Gamma_{\rm ej}\simeq\eta\,\frac{L_{\rm fun}}{\dot M_{\rm pair}c^2}\simeq\eta\times10,
\end{equation}
where $\eta<1$. A value of $\eta<0.5$ is favoured, inferred from the adiabatic expansion of supernova events \citep{sedov1959,ostrikermckee1988}. The factor of $\sim10$ follows from the energy partition of the wind injection. With $Y\simeq0.1$, the pair injection power is $\dot M_{\rm pair}c^2\simeq Y L_{\rm sd}$, and the numerical solution of Eq.\,\ref{eq:Efun} gives $L_{\rm fun}$ of the same order as $L_{\rm sd}$. 
For this mildly relativistic outflow, the motion is along the same direction as the GRB jet, but slower. The outflow trails the jet, and the interaction can occur only at late times, after the GRB jet has been substantially decelerated. Therefore, the early emission of the funnel outflow is dominated by the escaping photons,
\begin{equation}
\label{eq:Lob}
L_{\rm ob}\simeq(1-\eta)L_{\rm fun},
\end{equation}
and
\begin{equation}
\label{eq:Lobiso}
L_{\rm ob,iso}\simeq(1-\eta)\frac{L_{\rm fun}}{f_{\rm b}},
\end{equation}
for the isotropic-equivalent case.
Although the adiabatic expansion of the outflow itself consumes part of the radiation energy, the escaping photons still reflect the spectral state of the nebula fireball. The spectrum of the nebula is determined by the Compton $y$ parameter. For $y\gg1$ the scattering is saturated and the spectrum takes the Wien form, so the escaping photons carry a quasi-thermal spectrum; for $y\ll1$ photons escape before exchanging energy and the spectrum keeps the nonthermal injection shape. Figs.\,\ref{fig:fLPdV}, \ref{fig:fLPdV30}, and \ref{fig:ycompfunnel} all show the transition point for $y=1$. The funnel shifts the time at which $y$ reaches unity to the right by at most about $30\%$ compared to the no-funnel case of Fig.\,\ref{fig:ycomp}. Part of the fireball leaks through the funnel instead of driving the expansion, so the volume grows more slowly, which keeps $y$ above unity for longer. The parameter $y$ remains above unity at early times for the three magnetic fields and falls below unity on different timescales. For an observer, the thermal component could therefore appear in the extended emission of short GRBs. The peak photon energy is $3kT_{\rm C}$, from several keV to a few tens of keV depending on the magnetic field strength of the NS. As $y$ drops to unity the emission begins to convert into nonthermal. The spectrum then returns to that of the Metzger et al. models \citep{metzger2013supernova,metzgerpiro2014}, which lies in the soft X-ray band.

The funnel ultimately changes where the fireball energy goes. While the nebula remains optically thick, the funnel leakage is comparable to or exceeds the adiabatic loss. Since the emission is confined to the funnel opening angle, an observer within the beam infers an isotropic-equivalent luminosity much larger than the spin-down luminosity of the NS.

\section{Conclusion and Discussion}

In this paper we have investigated the evolution of the early nebula fireball powered by a post-merger NS. A post-merger NS with a high magnetic field injects a spin-down power into the surrounding ejecta \citep{zhangmeszaros2001}. The energy is deposited while the ejecta remain optically thick at early times, forming a radiation-dominated nebula. The cascade process injects copious $e^\pm$ pairs and keV photons into the nebula. The abundance of $e^\pm$ pairs keeps the mean free path of the photons far below the nebula radius. Since the Compton parameter $y>1$, the $e^\pm$ pairs and the photons are in saturated Comptonization and reach thermal equilibrium through Compton scattering, forming a so-called Wien fireball \citep{iwamototakahara2002,iwamototakahara2004}. If a jet-induced funnel persists, $e^\pm$ pairs and photons can leak out through the funnel, which dominates the early energy loss of the nebula fireball. At early times the funnel emission reflects the state of the nebula fireball, in particular the Compton temperature. For typical parameters the leaked photons peak at several keV to a few tens of keV. The nebula fireball can survive for more than $100\,\rm s$ depending on the magnetic field strength of the NS, and then the nebula converts to the normal nebula phase at $y=1$.

Quasi-thermal components with $kT\sim10$--$100$\,keV and durations of tens to hundreds of seconds are observed in the early-phase emission of a fraction of short GRBs \citep{chang2022,chang2024,peng2024}. A jet photosphere is usually assumed to account for the thermal component in prompt emission \citep{ryde2010}. However, sustaining a jet that emits steadily for tens to hundreds of seconds is difficult for a BNS merger, unless the merger involves a WD \citep{zhong2023nswd,chen2024nswd,wang2024nswd}. Instead, the nebula fireball is confined by the ejecta, and the ejecta expansion is slow enough to keep the Wien spectrum at keV energies for a longer time. We suggest that the quasi-thermal components observed in the extended emission of short GRBs come directly from the funnel emission of the Wien fireball. Several features distinguish this interpretation from the jet photosphere. The nebula fireball is in Compton equilibrium, so the thermal component should be fitted by a Wien spectrum rather than a blackbody. The energy injection from the central NS remains steady before the spin-down timescale, so the thermal component should accompany the plateau on light curves in observation. As the spin-down power declines and the Compton parameter falls below unity, the spectrum should change from thermal to nonthermal. The thermal component \citep{chang2022}, the plateau feature on the X-ray light curve \citep{sun2025}, and the change from a thermal to a synchrotron dominated spectrum reported for GRB\,230307A \citep{gupta2025} could support this suggestion.

A nebula fireball with a Wien spectrum does not necessarily require a merger, since any newborn millisecond NS can act as the central engine. A newborn NS can also form in massive-star collapses \citep{kasenbildsten2010}, in NS--WD mergers \citep{zhong2023nswd,chen2024nswd,wang2024nswd}, or in accretion-induced collapses of WDs \citep{yu2019aic,cheong2025aic}. Among these channels, BNS mergers have the smallest ejecta mass, of order $0.01\,M_\odot$ \citep{hotokezaka2013ejecta,radice2018,metzger2020kilonova}. A small ejecta mass helps the funnel to remain stably open for a longer time. Although the funnel forms while the jet propagates through the ejecta, continued injection from the NS is needed to counteract the refilling by the surrounding material \citep{wangzhangzhu2024}. For the collapse of a massive star, the prompt emission and the X-ray afterglow of the long GRB would likely overwhelm the funnel emission of the nebula fireball. Therefore, BNS mergers are the most promising case for observing the jet-induced funnel emission. In addition, in cocoon models \citep{nakarpiran2017,hamidani2023}, the nebula fireball can also power the cocoon through the funnel. The funnel can then be regarded as wrapped by the cocoon, and the energy that leaks through the funnel is absorbed by the cocoon and re-radiated.

\begin{acknowledgments}
This work is supported by the National Natural Science Foundation of China (grant No. 12303044 and 12393811).
\end{acknowledgments}

\bibliographystyle{aasjournalv7}
\bibliography{refs}

@ARTICLE{metzger2017welcome,
  author = {Metzger, Brian D.},
  title = {Welcome to the Multi-Messenger Era! Lessons from a Neutron Star Merger and the Landscape Ahead},
  journal = {arXiv e-prints},
  year = {2017},
  eprint = {1710.05931},
  archivePrefix = {arXiv},
}

@ARTICLE{metzger2013supernova,
  author = {Metzger, Brian D. and Vurm, Indrek and Hasco{\"e}t, Romain and Beloborodov, Andrei M.},
  title = {Ionization Break-out from Millisecond Pulsar Wind Nebulae: An X-ray Probe of the Origin of Superluminous Supernovae},
  journal = {MNRAS},
  year = {2013},
  volume = {437},
  pages = {703--720},
  doi = {10.1093/mnras/stt1922},
  eprint = {1307.8115},
  archivePrefix = {arXiv},
}

@ARTICLE{metzgerpiro2014,
  author       = {Metzger, B. D. and Piro, A. L.},
  title        = {Optical and X-ray Emission from Stable Millisecond Magnetars Formed from the Merger of Binary Neutron Stars},
  journal      = {MNRAS},
  year         = {2014},
  volume       = {439},
  pages        = {3916--3930},
  doi          = {10.1093/mnras/stu247},
  eprint       = {1311.1519},
  archivePrefix= {arXiv},
}

@BOOK{sedov1959,
  author       = {Sedov, L. I.},
  title        = {Similarity and Dimensional Methods in Mechanics},
  publisher    = {Academic Press},
  address      = {New York},
  year         = {1959},
}

@ARTICLE{ostrikermckee1988,
  author       = {Ostriker, J. P. and McKee, C. F.},
  title        = {Astrophysical blastwaves},
  journal      = {RvMP},
  year         = {1988},
  volume       = {60},
  pages        = {1--68},
  doi          = {10.1103/RevModPhys.60.1},
}

@ARTICLE{metzger2011proto,
  author       = {Metzger, B. D. and Giannios, D. and Thompson, T. A. and Bucciantini, N. and Quataert, E.},
  title        = {The Protomagnetar Model for Gamma-Ray Bursts},
  journal      = {MNRAS},
  year         = {2011},
  volume       = {413},
  pages        = {2031--2056},
  doi          = {10.1111/j.1365-2966.2011.18280.x},
  eprint       = {1012.0001},
  archivePrefix= {arXiv},
}

@ARTICLE{kasenbildsten2010,
  author       = {Kasen, D. and Bildsten, L.},
  title        = {Supernova Light Curves Powered by Young Magnetars},
  journal      = {ApJ},
  year         = {2010},
  volume       = {717},
  pages        = {245--249},
  doi          = {10.1088/0004-637X/717/1/245},
  eprint       = {0911.0680},
  archivePrefix= {arXiv},
}

@ARTICLE{yuzhanggao2013,
  author       = {Yu, Y.-W. and Zhang, B. and Gao, H.},
  title        = {Bright ``Merger-nova'' from the Remnant of a Neutron Star Binary Merger: A Signature of a Newly Born, Massive, Millisecond Magnetar},
  journal      = {ApJL},
  year         = {2013},
  volume       = {776},
  pages        = {L40},
  doi          = {10.1088/2041-8205/776/2/L40},
  eprint       = {1308.0876},
  archivePrefix= {arXiv},
}

@ARTICLE{yulidai2015,
  author       = {Yu, Y.-W. and Li, S.-Z. and Dai, Z.-G.},
  title        = {Rapidly Evolving and Luminous Transients Driven by Newly Born Neutron Stars},
  journal      = {ApJL},
  year         = {2015},
  volume       = {806},
  pages        = {L6},
  doi          = {10.1088/2041-8205/806/1/L6},
  eprint       = {1505.03251},
  archivePrefix= {arXiv},
}

@ARTICLE{yuliu2017,
  author       = {Yu, Y.-W. and Liu, L.-D. and Dai, Z.-G.},
  title        = {A Long-lived Remnant Neutron Star after GW170817 Inferred from Its Associated Kilonova},
  journal      = {ApJ},
  year         = {2018},
  volume       = {861},
  pages        = {114},
  doi          = {10.3847/1538-4357/aac6e5},
  eprint       = {1711.01898},
  archivePrefix= {arXiv},
}

@ARTICLE{aizhang2018,
  author       = {Ai, S. and Gao, H. and Dai, Z.-G. and Wu, X.-F. and Li, A. and Zhang, B.},
  title        = {The Allowed Parameter Space of a Long-lived Neutron Star as the Merger Remnant of GW170817},
  journal      = {ApJ},
  year         = {2018},
  volume       = {860},
  pages        = {57},
  doi          = {10.3847/1538-4357/aac2b7},
  eprint       = {1802.00571},
  archivePrefix= {arXiv},
}

@ARTICLE{svensson1987,
  author       = {Svensson, R.},
  title        = {Non-thermal Pair Production in Compact X-ray Sources: First-Order Compton Cascades in Soft Radiation Fields},
  journal      = {MNRAS},
  year         = {1987},
  volume       = {227},
  pages        = {403--451},
  doi          = {10.1093/mnras/227.2.403},
}

@ARTICLE{goldreichjulian1969,
  author       = {Goldreich, P. and Julian, W. H.},
  title        = {Pulsar Electrodynamics},
  journal      = {ApJ},
  year         = {1969},
  volume       = {157},
  pages        = {869--880},
  doi          = {10.1086/150119},
}

@ARTICLE{iwamototakahara2002,
  author       = {Iwamoto, S. and Takahara, F.},
  title        = {Relativistic Outflow of Electron-Positron Pair Plasma from a Wien Equilibrium State},
  journal      = {ApJ},
  year         = {2002},
  volume       = {565},
  pages        = {163--173},
  doi          = {10.1086/324480},
  eprint       = {astro-ph/0204223},
  archivePrefix= {arXiv},
}

@ARTICLE{iwamototakahara2004,
  author       = {Iwamoto, S. and Takahara, F.},
  title        = {Wien Fireball Model of Relativistic Outflows in Active Galactic Nuclei},
  journal      = {ApJ},
  year         = {2004},
  volume       = {601},
  pages        = {78--89},
  doi          = {10.1086/380300},
  eprint       = {astro-ph/0309782},
  archivePrefix= {arXiv},
}

@ARTICLE{zhang2023compton,
  author       = {Zhang, Z. J. and Zhang, B.-B. and Meng, Y.-Z.},
  title        = {A Comptonized Fireball Bubble: Physical Origin of Magnetar Giant Flares},
  journal      = {MNRAS},
  year         = {2023},
  volume       = {520},
  pages        = {6195--6213},
  doi          = {10.1093/mnras/stad443},
  eprint       = {2109.14252},
  archivePrefix= {arXiv},
}

@ARTICLE{yin2024compton,
  author       = {Yin, Y.-H. I. and Zhang, Z. J. and Yang, J. and Chen, R.-C. and Rehman, U. and Varun and Zhang, B.-B. and others},
  title        = {A Comptonized Fireball Bubble Fits the Second Extragalactic Magnetar Giant Flare GRB 231115A},
  journal      = {ApJL},
  year         = {2024},
  volume       = {963},
  pages        = {L10},
  doi          = {10.3847/2041-8213/ad2839},
  eprint       = {2312.14833},
  archivePrefix= {arXiv},
}

@ARTICLE{price2006,
  author = {Price, D. J. and Rosswog, S.},
  title = {Producing Ultrastrong Magnetic Fields in Neutron Star Mergers},
  journal = {Science},
  year = {2006},
  volume = {312},
  pages = {719--722},
  doi = {10.1126/science.1125201},
  eprint = {astro-ph/0603845},
  archivePrefix = {arXiv},
}

@ARTICLE{zrake2013,
  author = {Zrake, J. and MacFadyen, A. I.},
  title = {Magnetic Energy Production by Turbulence in Binary Neutron Star Mergers},
  journal = {ApJL},
  year = {2013},
  volume = {769},
  pages = {L29},
  doi = {10.1088/2041-8205/769/2/L29},
  eprint = {1303.1450},
  archivePrefix = {arXiv},
}

@ARTICLE{kiuchi2024,
  author = {Kiuchi, K. and others},
  title = {A Large-scale Magnetic Field Produced by a Solar-like Dynamo in Binary Neutron Star Mergers},
  journal = {NatAs},
  year = {2024},
  volume = {8},
  pages = {298--307},
  doi = {10.1038/s41550-024-02194-y},
  eprint = {2306.15721},
  archivePrefix = {arXiv},
}

@ARTICLE{sarin2021,
  author = {Sarin, N. and Lasky, P. D.},
  title = {The Evolution of Binary Neutron Star Post-merger Remnants: A Review},
  journal = {GReGr},
  year = {2021},
  volume = {53},
  pages = {59},
  doi = {10.1007/s10714-021-02831-1},
  eprint = {2012.08172},
  archivePrefix = {arXiv},
}

@ARTICLE{lightmanzdziarski1987,
  author = {Lightman, A. P. and Zdziarski, A. A.},
  title = {Pair Production and Compton Scattering in Compact Sources and Comparison to Observations of Active Galactic Nuclei},
  journal = {ApJ},
  year = {1987},
  volume = {319},
  pages = {643},
  doi = {10.1086/165485},
}

@ARTICLE{vurmpoutanen2009,
  author = {Vurm, I. and Poutanen, J.},
  title = {Time-dependent Modeling of Radiative Processes in Hot Magnetized Plasmas},
  journal = {ApJ},
  year = {2009},
  volume = {698},
  pages = {293--316},
  doi = {10.1088/0004-637X/698/1/293},
}

@ARTICLE{nakarpiran2017,
  author       = {Nakar, E. and Piran, T.},
  title        = {The Observable Signatures of GRB Cocoons},
  journal      = {ApJ},
  year         = {2017},
  volume       = {834},
  pages        = {28},
  doi          = {10.3847/1538-4357/834/1/28},
  eprint       = {1610.05362},
  archivePrefix= {arXiv},
}

@ARTICLE{hamidani2023,
  author       = {Hamidani, H. and Ioka, K.},
  title        = {Cocoon breakout and escape from the ejecta of neutron star mergers},
  journal      = {MNRAS},
  year         = {2023},
  volume       = {520},
  pages        = {1111--1127},
  doi          = {10.1093/mnras/stad041},
  eprint       = {2210.00814},
  archivePrefix= {arXiv},
}

@ARTICLE{cheong2025aic,
  author       = {Cheong, P. C.-K. and Pitik, T. and Longo Micchi, L. F. and Radice, D.},
  title        = {Gamma-Ray Bursts and Kilonovae from the Accretion-induced Collapse of White Dwarfs},
  journal      = {ApJL},
  year         = {2025},
  volume       = {978},
  pages        = {L38},
  doi          = {10.3847/2041-8213/ada1cc},
  eprint       = {2410.10938},
  archivePrefix= {arXiv},
}

@ARTICLE{yu2019aic,
  author       = {Yu, Y.-W. and Chen, A. and Li, X.-D.},
  title        = {X-Ray Transients from the Accretion-induced Collapse of White Dwarfs},
  journal      = {ApJL},
  year         = {2019},
  volume       = {877},
  pages        = {L21},
  doi          = {10.3847/2041-8213/ab1f85},
  eprint       = {1904.11840},
  archivePrefix= {arXiv},
}

@ARTICLE{wangzhangzhu2024,
  author       = {Wang, Y.-H. and Zhang, B. and Zhu, Z.},
  title        = {Anisotropic Energy Injection from Magnetar Central Engines in Short GRBs},
  journal      = {MNRAS},
  year         = {2024},
  volume       = {528},
  pages        = {3705--3718},
  doi          = {10.1093/mnras/stae136},
  eprint       = {2309.15141},
  archivePrefix= {arXiv},
}

@ARTICLE{radice2016,
  author = {Radice, David and Galeazzi, Filippo and Lippuner, Jonas and Roberts, Luke F. and Ott, Christian D. and Rezzolla, Luciano},
  title = {Dynamical Mass Ejection from Binary Neutron Star Mergers},
  journal = {MNRAS},
  year = {2016},
  volume = {460},
  pages = {3255--3271},
  doi = {10.1093/mnras/stw1227},
  eprint = {1601.02426},
  archivePrefix = {arXiv},
}

@ARTICLE{hotokezaka2013ejecta,
  author = {Hotokezaka, Kenta and Kiuchi, Kenta and Kyutoku, Koutarou and Okawa, Hirotada and Sekiguchi, Yu-ichiro and Shibata, Masaru},
  title = {Mass Ejection from the Merger of Binary Neutron Stars},
  journal = {PhRvD},
  year = {2013},
  volume = {87},
  pages = {024001},
  doi = {10.1103/PhysRevD.87.024001},
  eprint = {1212.0905},
  archivePrefix = {arXiv},
}

@ARTICLE{shibatahotokezaka2019,
  author = {Shibata, Masaru and Hotokezaka, Kenta},
  title = {Merger and Mass Ejection of Neutron-Star Binaries},
  journal = {ARNPS},
  year = {2019},
  volume = {69},
  pages = {41--64},
  doi = {10.1146/annurev-nucl-101918-023625},
  eprint = {1908.02350},
  archivePrefix = {arXiv},
}

@ARTICLE{abbott2017,
  author = {Abbott, B. P. and others},
  title = {Multi-messenger Observations of a Binary Neutron Star Merger},
  journal = {ApJL},
  year = {2017},
  volume = {848},
  pages = {L12},
  doi = {10.3847/2041-8213/aa91c9},
  eprint = {1710.05833},
  archivePrefix = {arXiv},
}

@ARTICLE{cowperthwaite2017,
  author = {Cowperthwaite, P. S. and Berger, E. and Villar, V. A. and Metzger, B. D. and Nicholl, M. and Chornock, R. and Blanchard, P. K. and Fong, W. and Margutti, R. and Soares-Santos, M.},
  title = {The Electromagnetic Counterpart of the Binary Neutron Star Merger LIGO/Virgo GW170817. II. UV, Optical, and Near-infrared Light Curves and Comparison to Kilonova Models},
  journal = {ApJ},
  year = {2017},
  volume = {848},
  pages = {L17},
  doi = {10.3847/2041-8213/aa8fc7},
  eprint = {1710.05840},
  archivePrefix = {arXiv},
}

@ARTICLE{drout2017,
  author = {Drout, M. R. and Piro, A. L. and Shappee, B. J. and Kilpatrick, C. D. and Simon, J. D. and Contreras, C. and Coulter, D. A. and Foley, R. J. and Siebert, M. R. and Morrell, N.},
  title = {Light Curves of the Neutron Star Merger GW170817/SSS17a: Implications for R-Process Nucleosynthesis},
  journal = {Science},
  year = {2017},
  volume = {358},
  pages = {1570--1574},
  doi = {10.1126/science.aaq0049},
  eprint = {1710.05443},
  archivePrefix = {arXiv},
}

@ARTICLE{tanvir2017,
  author = {Tanvir, N. R. and Levan, A. J. and Gonzalez-Fernandez, C. and Korobkin, O. and Mandel, I. and Rosswog, S. and Hjorth, J. and D'Avanzo, P. and Fruchter, A. S. and Fryer, C. L.},
  title = {The Emergence of a Lanthanide-Rich Kilonova Following the Merger of Two Neutron Stars},
  journal = {ApJL},
  year = {2017},
  volume = {848},
  pages = {L27},
  doi = {10.3847/2041-8213/aa90b6},
  eprint = {1710.05455},
  archivePrefix = {arXiv},
}

@ARTICLE{metzger2020kilonova,
  author = {Metzger, Brian D.},
  title = {Kilonovae},
  journal = {LRR},
  year = {2020},
  volume = {23},
  pages = {1},
  doi = {10.1007/s41114-019-0024-0},
  eprint = {1910.01617},
  archivePrefix = {arXiv},
}

@ARTICLE{gompertz2018diversity,
  author = {Gompertz, B. P. and Levan, A. J. and Tanvir, N. R. and Hjorth, J. and Covino, S. and Evans, P. A. and Fruchter, A. S. and Gonz{\'a}lez-Fern{\'a}ndez, C. and Jin, Z. P. and Lyman, J. D. and Oates, S. R. and O'Brien, P. T. and Wiersema, K.},
  title = {The Diversity of Kilonova Emission in Short Gamma-Ray Bursts},
  journal = {ApJ},
  year = {2018},
  volume = {860},
  pages = {62},
  doi = {10.3847/1538-4357/aac206},
  eprint = {1710.05442},
  archivePrefix = {arXiv},
}

@ARTICLE{kawaguchi2020diversity,
  author = {Kawaguchi, Kyohei and Shibata, Masaru and Tanaka, Masaomi},
  title = {Diversity of Kilonova Light Curves},
  journal = {ApJ},
  year = {2020},
  volume = {889},
  pages = {171},
  doi = {10.3847/1538-4357/ab61f6},
  eprint = {1908.05815},
  archivePrefix = {arXiv},
}

@ARTICLE{margalitmetzger2017,
  author = {Margalit, Ben and Metzger, Brian},
  title = {Constraining the Maximum Mass of Neutron Stars From Multi-Messenger Observations of GW170817},
  journal = {ApJL},
  year = {2017},
  volume = {850},
  pages = {L19},
  doi = {10.3847/2041-8213/aa991c},
  eprint = {1710.05938},
  archivePrefix = {arXiv},
}

@ARTICLE{rezzolla2018,
  author = {Rezzolla, Luciano and Most, Elias R. and Weih, Lukas R.},
  title = {Using gravitational-wave observations and quasi-universal relations to constrain the maximum mass of neutron stars},
  journal = {ApJL},
  year = {2018},
  volume = {852},
  pages = {L25},
  doi = {10.3847/2041-8213/aaa401},
  eprint = {1711.00314},
  archivePrefix = {arXiv},
}

@ARTICLE{radice2018,
  author = {Radice, David and Perego, Albino and Zappa, Francesco and Bernuzzi, Sebastiano},
  title = {GW170817: Joint Constraint on the Neutron Star Equation of State from Multimessenger Observations},
  journal = {ApJL},
  year = {2018},
  volume = {852},
  pages = {L29},
  doi = {10.3847/2041-8213/aaa402},
  eprint = {1711.03647},
  archivePrefix = {arXiv},
}

@ARTICLE{radicelonglived2018,
  author       = {Radice, David and Perego, Albino and Bernuzzi, Sebastiano and Zhang, Bing},
  title        = {Long-lived Remnants from Binary Neutron Star Mergers},
  journal      = {MNRAS},
  year         = {2018},
  volume       = {481},
  pages        = {3670--3682},
  doi          = {10.1093/mnras/sty2531},
  eprint       = {1803.10865},
  archivePrefix= {arXiv},
}

@ARTICLE{lu2015,
  author = {L{\"u}, Hou-Jun and Zhang, Bing and Lei, Wei-Hua and Li, Ye and Lasky, Paul D},
  title = {The Millisecond Magnetar Central Engine in short GRBs},
  journal = {ApJ},
  year = {2015},
  volume = {805},
  pages = {89},
  doi = {10.1088/0004-637X/805/2/89},
  eprint = {1501.02589},
  archivePrefix = {arXiv},
}

@ARTICLE{liliuyuzhang2018,
  author       = {Li, Shao-Ze and Liu, Liang-Duan and Yu, Yun-Wei and Zhang, Bing},
  title        = {What Powered the Optical Transient AT2017gfo Associated with GW170817?},
  journal      = {ApJL},
  year         = {2018},
  volume       = {861},
  pages        = {L12},
  doi          = {10.3847/2041-8213/aace61},
  eprint       = {1804.06597},
  archivePrefix= {arXiv},
}

@ARTICLE{stratta2018,
  author = {Stratta, G. and Dainotti, M. G. and Dall'Osso, S. and Hernandez, X. and De Cesare, G.},
  title = {On the magnetar origin of the GRBs presenting X-ray afterglow plateaus},
  journal = {ApJ},
  year = {2018},
  volume = {869},
  pages = {155},
  doi = {10.3847/1538-4357/aadd8f},
  eprint = {1804.08652},
  archivePrefix = {arXiv},
}

@ARTICLE{gibson2017,
  author = {Gibson, Sarah and Wynn, Graham and Gompertz, Benjamin and O'Brien, Paul},
  title = {Fallback Accretion on to a Newborn Magnetar: Short GRBs with Extended Emission},
  journal = {MNRAS},
  year = {2017},
  volume = {470},
  pages = {4925--4940},
  doi = {10.1093/mnras/stx1531},
  eprint = {1706.04802},
  archivePrefix = {arXiv},
}

@ARTICLE{mosta2020,
  author = {M{\"o}sta, Philipp and Radice, David and Haas, Roland and Schnetter, Erik and Bernuzzi, Sebastiano},
  title = {A magnetar engine for short GRBs and kilonovae},
  journal = {ApJL},
  year = {2020},
  volume = {901},
  pages = {L37},
  doi = {10.3847/2041-8213/abb6ef},
  eprint = {2003.06043},
  archivePrefix = {arXiv},
}

@ARTICLE{gao2015mergernova,
  author = {Gao, He and Ding, Xuan and Wu, Xue-Feng and Dai, Zi-Gao and Zhang, Bing},
  title = {GRB 080503 late afterglow re-brightening: signature of a magnetar powered merger-nova},
  journal = {ApJ},
  year = {2015},
  volume = {807},
  pages = {163},
  doi = {10.1088/0004-637X/807/2/163},
  eprint = {1506.06816},
  archivePrefix = {arXiv},
}

@ARTICLE{sun2025,
  author = {Sun, H. and Wang, C.-W. and Yang, J. and Zhang, B.-B. and Xiong, S.-L. and Yin, Y.-H. I. and Liu, Y. and Li, Y. and Xue, W.-C. and Yan, Z.-Y.},
  title = {Magnetar emergence in a peculiar gamma-ray burst from a compact star merger},
  journal = {NSR},
  year = {2025},
  volume = {12},
  pages = {nwae401},
  doi = {10.1093/nsr/nwae401},
  eprint = {2307.05689},
  archivePrefix = {arXiv},
}

@ARTICLE{zhong2024,
  author = {Zhong, Shu-Qing and Li, Long and Xiao, Di and Sun, Hui and Zhang, Bin-Bin and Dai, Zi-Gao},
  title = {The Very Early Soft X-ray Plateau of GRB 230307A: Signature of an Evolving Radiative Efficiency in Magnetar Wind Dissipation?},
  journal = {ApJL},
  year = {2024},
  volume = {963},
  pages = {L26},
  doi = {10.3847/2041-8213/ad2852},
  eprint = {2402.10619},
  archivePrefix = {arXiv},
}

@ARTICLE{fraija2026,
  author = {Fraija, Nissim},
  title = {Long-Duration GRB 211211A: Internal Energy Dissipation Driven by a Long-Lived Magnetar},
  journal = {MNRAS},
  year = {2026},
  volume = {549},
  pages = {stag949},
  doi = {10.1093/mnras/stag949},
  eprint = {2605.30025},
  archivePrefix = {arXiv},
}

@ARTICLE{chen2025,
  author = {Chen, R.-C. and Zhang, B. and Xiong, S.-L. and Zhang, B.-B. and others},
  title = {Evidence for a Brief Appearance of Gamma-ray Periodicity after a Compact Star Merger},
  journal = {NatAs},
  year = {2025},
  volume = {9},
  pages = {1701--1713},
  doi = {10.1038/s41550-025-02649-w},
  eprint = {2509.15824},
  archivePrefix = {arXiv},
}

@ARTICLE{chang2024,
  author = {Chang, Xue-Zhao and L{\"u}, Hou-Jun and Yang, Xing and Chen, Jia-Ming and Liang, En-Wei},
  title = {The Thermal Emission in Short Gamma-Ray Bursts with Extended Emission Observed by Fermi/GBM},
  journal = {ApJS},
  year = {2024},
  volume = {275},
  pages = {9},
  doi = {10.3847/1538-4365/ad7eaf},
  eprint = {2410.09415},
  archivePrefix = {arXiv},
}

@ARTICLE{wang2023photosphere,
  author = {Wang, Yun and Xia, Zi-Qing and Zheng, Tian-Ci and Ren, Jia and Fan, Yi-Zhong},
  title = {A Broken ``$\alpha$-intensity'' Relation Caused by the Evolving Photosphere Emission and the Nature of the Extraordinarily Bright GRB 230307A},
  journal = {ApJL},
  year = {2023},
  volume = {953},
  pages = {L8},
  doi = {10.3847/2041-8213/ace7d4},
  eprint = {2303.11083},
  archivePrefix = {arXiv},
}

@ARTICLE{gupta2025,
  author = {Gupta, Soumya and Gupta, Rahul and Chattopadhayay, Tanmoy and Sahayanathan, Sunder and Frederiks, D. and Svinkin, D. and Bhattacharya, Dipankar and Racusin, Judith and Vadawale, Santosh and Bhalerao, Varun},
  title = {Time-resolved Spectro-polarimetric Analysis of Extremely Bright GRB 230307A: Possible Evidence of Evolution from Photospheric to Synchrotron Dominated Emission},
  journal = {A\&A},
  year = {2025},
  volume = {701},
  pages = {A172},
  doi = {10.1051/0004-6361/202555055},
  eprint = {2504.05038},
  archivePrefix = {arXiv},
}

@ARTICLE{wistemar2025,
  author = {Wistemar, Oscar and Alamaa, Filip and Ryde, Felix},
  title = {Photospheric Emission from GRB 211211A Altered by a Strong Radiation-mediated Shock},
  journal = {MNRAS},
  year = {2025},
  volume = {544},
  pages = {3683--3695},
  doi = {10.1093/mnras/staf1757},
  eprint = {2506.08122},
  archivePrefix = {arXiv},
}

@ARTICLE{rastinejad2022,
  author = {Rastinejad, J. C. and Gompertz, B. P. and Levan, A. J. and Fong, W. and Nicholl, M. and Lamb, G. P. and Nugent, Anya E. and Oates, Samantha R. and Tanvir, Nial R. and de Ugarte Postigo, Antonio},
  title = {A Kilonova Following a Long-Duration Gamma-Ray Burst at 350 Mpc},
  journal = {Nature},
  year = {2022},
  volume = {612},
  pages = {223--227},
  doi = {10.1038/s41586-022-05390-w},
  eprint = {2204.10864},
  archivePrefix = {arXiv},
}

@ARTICLE{chang2022,
  author = {Chang, Xue-Zhao and L{\"u}, Hou-Jun and Yang, Xing and Chen, Jia-Ming and Liang, En-Wei},
  title = {Thermal and nonthermal emission from a peculiar long-duration GRB 211211A},
  journal = {ApJ},
  year = {2023},
  volume = {943},
  pages = {146},
  doi = {10.3847/1538-4357/aca969},
  eprint = {2206.11438},
  archivePrefix = {arXiv},
}

@ARTICLE{yang2024,
  author = {Yang, Yu-Han and Troja, Eleonora and O'Connor, Brendan and Fryer, Chris L. and Im, Myungshin and Durbak, Joe and Paek, Gregory S. H. and Ricci, Roberto and De Bom, Clécio R. and Gillanders, James H.},
  title = {A lanthanide-rich kilonova in the aftermath of a long gamma-ray burst},
  journal = {Nature},
  year = {2024},
  volume = {626},
  pages = {742--745},
  doi = {10.1038/s41586-023-06979-5},
  eprint = {2308.00638},
  archivePrefix = {arXiv},
}

@ARTICLE{peng2024,
  author = {Peng, Zhao-Yang and Chen, Jia-Ming and Mao, Jirong},
  title = {A Comparative Analysis of Two Peculiar Gamma-Ray Bursts: GRB 230307A and GRB 211211A},
  journal = {ApJ},
  year = {2024},
  volume = {969},
  pages = {26},
  doi = {10.3847/1538-4357/ad45fc},
  eprint = {2404.17913},
  archivePrefix = {arXiv},
}

@ARTICLE{dailu1998a,
  author       = {Dai, Zi-Gao and Lu, Tan},
  title        = {Gamma-Ray Burst Afterglows and Evolution of Postburst Fireballs with Energy Injection from Strongly Magnetic Millisecond Pulsars},
  journal      = {A\&A},
  year         = {1998},
  volume       = {333},
  pages        = {L87--L90},
  eprint       = {astro-ph/9810402},
  archivePrefix= {arXiv},
}

@ARTICLE{dailu1998b,
  author       = {Dai, Zi-Gao and Lu, Tan},
  title        = {Gamma-Ray Bursts and Afterglows from Rotating Strange Stars and Neutron Stars},
  journal      = {PhRvL},
  year         = {1998},
  volume       = {81},
  pages        = {4301--4304},
  doi          = {10.1103/PhysRevLett.81.4301},
}

@ARTICLE{zhangmeszaros2001,
  author       = {Zhang, Bing and M{\'e}sz{\'a}ros, Peter},
  title        = {Gamma-Ray Burst Afterglow with Continuous Energy Injection: Signature of a Highly Magnetized Millisecond Pulsar},
  journal      = {ApJ},
  year         = {2001},
  volume       = {552},
  pages        = {L35--L38},
  doi          = {10.1086/320255},
  eprint       = {astro-ph/0011133},
  archivePrefix= {arXiv},
}

@ARTICLE{lipaczynski1998,
  author       = {Li, Li-Xin and Paczynski, Bohdan},
  title        = {Transient Events from Neutron Star Mergers},
  journal      = {ApJ},
  year         = {1998},
  volume       = {507},
  pages        = {L59--L62},
  doi          = {10.1086/311680},
  eprint       = {astro-ph/9807272},
  archivePrefix= {arXiv},
}

@ARTICLE{metzger2010,
  author       = {Metzger, B. D. and Mart{\'i}nez-Pinedo, G. and Darbha, S. and Quataert, E. and Arcones, A. and Kasen, D. and Thomas, R. and Nugent, P. and Panov, I. V. and Zinner, N. T.},
  title        = {Electromagnetic Counterparts of Compact Object Mergers Powered by the Radioactive Decay of R-Process Nuclei},
  journal      = {MNRAS},
  year         = {2010},
  volume       = {406},
  pages        = {2650--2662},
  doi          = {10.1111/j.1365-2966.2010.16864.x},
  eprint       = {1001.5029},
  archivePrefix= {arXiv},
}

@ARTICLE{tanvir2013,
  author       = {Tanvir, N. R. and Levan, A. J. and Fruchter, A. S. and Hjorth, J. and Hounsell, R. A. and Wiersema, K. and Tunnicliffe, R. L.},
  title        = {A ``Kilonova'' Associated with the Short-Duration $\gamma$-Ray Burst GRB 130603B},
  journal      = {Nature},
  year         = {2013},
  volume       = {500},
  pages        = {547--549},
  doi          = {10.1038/nature12505},
  eprint       = {1306.4971},
  archivePrefix= {arXiv},
}

@ARTICLE{berger2013,
  author       = {Berger, E. and Fong, W. and Chornock, R.},
  title        = {An R-Process Kilonova Associated with the Short-Hard GRB 130603B},
  journal      = {ApJL},
  year         = {2013},
  volume       = {774},
  pages        = {L23},
  doi          = {10.1088/2041-8205/774/2/L23},
  eprint       = {1306.3960},
  archivePrefix= {arXiv},
}

@ARTICLE{zhang2013,
  author       = {Zhang, Bing},
  title        = {Early X-Ray and Optical Afterglow of Gravitational Wave Bursts from Mergers of Binary Neutron Stars},
  journal      = {ApJL},
  year         = {2013},
  volume       = {763},
  pages        = {L22},
  doi          = {10.1088/2041-8205/763/1/L22},
  eprint       = {1212.0773},
  archivePrefix= {arXiv},
}

@ARTICLE{bucciantini2009,
  author       = {Bucciantini, N. and Quataert, E. and Metzger, B. D. and Thompson, T. A. and Arons, J. and Del Zanna, L.},
  title        = {Magnetized Relativistic Jets and Long-Duration GRBs from Magnetar Spin-Down during Core-Collapse Supernovae},
  journal      = {MNRAS},
  year         = {2009},
  volume       = {396},
  pages        = {2038--2050},
  doi          = {10.1111/j.1365-2966.2009.14940.x},
  eprint       = {0901.3801},
  archivePrefix= {arXiv},
}

@ARTICLE{bucciantini2012,
  author       = {Bucciantini, N. and Metzger, B. D. and Thompson, T. A. and Quataert, E.},
  title        = {Short Gamma-Ray Bursts with Extended Emission from Magnetar Birth: Jet Formation and Collimation},
  journal      = {MNRAS},
  year         = {2012},
  volume       = {419},
  pages        = {1537--1545},
  doi          = {10.1111/j.1365-2966.2011.19810.x},
  eprint       = {1106.4668},
  archivePrefix= {arXiv},
}

@ARTICLE{gao2017mergernova,
  author       = {Gao, He and Zhang, Bing and L{\"u}, Hou-Jun and Li, Ye},
  title        = {Searching for Magnetar-powered Merger-novae from Short GRBs},
  journal      = {ApJ},
  year         = {2017},
  volume       = {837},
  pages        = {50},
  doi          = {10.3847/1538-4357/aa5be3},
  eprint       = {1608.03375},
  archivePrefix= {arXiv},
}

@ARTICLE{fan2013magnetar,
  author       = {Fan, Y.-Z. and Yu, Y.-W. and Xu, D. and Jin, Z.-P. and Wu, X.-F. and Wei, D.-M. and Zhang, B.},
  title        = {A Supramassive Magnetar Central Engine for GRB 130603B},
  journal      = {ApJL},
  year         = {2013},
  volume       = {779},
  pages        = {L25},
  doi          = {10.1088/2041-8205/779/2/L25},
  eprint       = {1311.7185},
  archivePrefix= {arXiv},
}

@ARTICLE{zhong2023nswd,
  author = {Zhong, Shu-Qing and Li, Long and Dai, Zi-Gao},
  title = {GRB 211211A: A Neutron Star--White Dwarf Merger?},
  journal = {ApJL},
  year = {2023},
  volume = {947},
  pages = {L21},
  doi = {10.3847/2041-8213/acca83},
  eprint = {2304.04009},
  archivePrefix = {arXiv},
}

@ARTICLE{wang2024nswd,
  author = {Wang, Xiangyu Ivy and Yu, Yun-Wei and Ren, Jia and Yang, Jun and Zou, Ze-Cheng and Zhu, Jin-Ping},
  title = {What Powered the Kilonova-like Emission after GRB 230307A in the Framework of a Neutron Star--White Dwarf Merger?},
  journal = {ApJL},
  year = {2024},
  volume = {964},
  pages = {L9},
  doi = {10.3847/2041-8213/ad2df6},
  eprint = {2402.11304},
  archivePrefix = {arXiv},
}

@ARTICLE{chen2024nswd,
  author = {Chen, Junping and Shen, Rong-Feng and Tan, Wen-Jun and Wang, Chen-Wei and Xiong, Shao-Lin and Chen, Run-Chao and Zhang, Bin-Bin},
  title = {Repeated Partial Disruptions in a White Dwarf--Neutron Star or White Dwarf--Black Hole Merger Modulate the Prompt Emission of Long-duration Merger-type GRBs},
  journal = {ApJL},
  year = {2024},
  volume = {973},
  pages = {L33},
  doi = {10.3847/2041-8213/ad7737},
  eprint = {2409.00472},
  archivePrefix = {arXiv},
}

@ARTICLE{shibatazhou2019,
  author = {Shibata, Masaru and Zhou, Enping and Kiuchi, Kenta and Fujibayashi, Sho},
  title = {Constraint on the Maximum Mass of Neutron Stars Using GW170817 Event},
  journal = {PhRvD},
  year = {2019},
  volume = {100},
  pages = {023015},
  doi = {10.1103/PhysRevD.100.023015},
  eprint = {1905.03656},
  archivePrefix = {arXiv},
}

@ARTICLE{ai2020,
  author = {Ai, Shunke and Gao, He and Zhang, Bing},
  title = {What Constraints on the Neutron Star Maximum Mass Can One Pose from GW170817 Observations?},
  journal = {ApJ},
  year = {2020},
  volume = {893},
  pages = {146},
  doi = {10.3847/1538-4357/ab80bd},
  eprint = {1912.06369},
  archivePrefix = {arXiv},
}

@ARTICLE{gao2022,
  author = {Gao, He and Lei, Wei-Hua and Zhu, Zi-Pei},
  title = {GRB 211211A: A Prolonged Central Engine under a Strong Magnetic Field Environment},
  journal = {ApJL},
  year = {2022},
  volume = {934},
  pages = {L12},
  doi = {10.3847/2041-8213/ac80c7},
  eprint = {2205.05031},
  archivePrefix = {arXiv},
}

@ARTICLE{zhang2025,
  author = {Zhang, Bing},
  title = {On the Duration of Gamma-Ray Bursts},
  journal = {JHEAp},
  year = {2025},
  volume = {45},
  pages = {325--332},
  doi = {10.1016/j.jheap.2024.12.013},
  eprint = {2501.00239},
  archivePrefix = {arXiv},
}

@ARTICLE{gao2013afterglow,
  author = {Gao, He and Ding, Xuan and Wu, Xue-Feng and Zhang, Bing and Dai, Zi-Gao},
  title = {Bright Broadband Afterglows of Gravitational Wave Bursts from Mergers of Binary Neutron Stars},
  journal = {ApJ},
  year = {2013},
  volume = {771},
  pages = {86},
  doi = {10.1088/0004-637X/771/2/86},
  eprint = {1301.0439},
  archivePrefix = {arXiv},
}

@ARTICLE{ryde2010,
  author = {Ryde, F. and Axelsson, M. and Zhang, B. B. and McGlynn, S. and Pe'er, A. and Lundman, C. and Larsson, S. and Battelino, M. and Zhang, B. and Bissaldi, E. and others},
  title = {Identification and Properties of the Photospheric Emission in GRB090902B},
  journal = {ApJL},
  year = {2010},
  volume = {709},
  pages = {L172--L177},
  doi = {10.1088/2041-8205/709/2/L172},
  eprint = {0911.2025},
  archivePrefix = {arXiv},
}

@ARTICLE{begelman1989,
  author = {Begelman, M. C. and Cioffi, D. F.},
  title = {Overpressured Cocoons in Extragalactic Radio Sources},
  journal = {ApJ},
  year = {1989},
  volume = {345},
  pages = {L21},
  doi = {10.1086/185542},
}

@ARTICLE{irwin2019,
  author = {Irwin, C. M. and Nakar, E. and Piran, T.},
  title = {The Propagation of Choked Jet Outflows in Power-Law External Media},
  journal = {MNRAS},
  year = {2019},
  volume = {489},
  pages = {2844--2872},
  doi = {10.1093/mnras/stz2268},
  eprint = {1907.04985},
  archivePrefix = {arXiv},
}

@ARTICLE{sekiguchi2015,
  author = {Sekiguchi, Y. and Kiuchi, K. and Kyutoku, K. and Shibata, M.},
  title = {Dynamical Mass Ejection from Binary Neutron Star Mergers: Radiation-Hydrodynamics Study in General Relativity},
  journal = {PhRvD},
  year = {2015},
  volume = {91},
  pages = {064059},
  doi = {10.1103/PhysRevD.91.064059},
  eprint = {1502.06660},
  archivePrefix = {arXiv},
}

@ARTICLE{yang2022,
  author = {Yang, Jun and Ai, Shunke and Zhang, Bin-Bin and Zhang, Bing and Liu, Zi-Ke and Wang, Xiangyu Ivy and Yang, Yu-Han and Yin, Yi-Han and Li, Ye and L{\"u}, Hou-Jun},
  title = {A Long-Duration Gamma-Ray Burst with a Peculiar Origin},
  journal = {Nature},
  year = {2022},
  volume = {612},
  pages = {232--235},
  doi = {10.1038/s41586-022-05403-8},
  eprint = {2204.12771},
  archivePrefix = {arXiv},
}

\end{CJK*}
\end{document}